\documentclass{aa}  

\usepackage{graphicx}
\usepackage{subfig}
\usepackage{xcolor}
\usepackage{txfonts}
\begin{document}

   \title{Exploring compact binary populations with the Einstein Telescope}

   %\subtitle{}

   \author{Neha Singh,
          \inst{1,}
          \inst{2}\thanks{\email{singh@lapth.cnrs.fr}}%\thanks{\email{nsingh@astrouw.edu.pl}}
          %\and
          Tomasz Bulik,
          \inst{2}
          Krzysztof Belczynski,
          \inst{3}
          %M. Giersz
          %\inst{2}
          Abbas Askar
          \inst{4}
          }
   \institute{Laboratoire d’Annecy-le-Vieux de Physique Théorique (LAPTh), USMB, CNRS, F-74940 Annecy, France. \and
             Astronomical Observatory, University of Warsaw, Al. Ujazdowskie 4, 00-478 Warsaw, Poland. \and  
             Nicolaus Copernicus Astronomical Center, Polish Academy of Sciences, ul. Bartycka 18, 00-716 Warsaw, Poland. \and
             Lund Observatory, Department of Astronomy, and Theoretical Physics, Lund University, Box 43, SE-221 00 Lund, Sweden.
             }
    \authorrunning {N. Singh, T. Bulik, K. Belczynski, A. Askar}

   \date{Received ; accepted  }

% \abstract{}{}{}{}{} 
% 5 {} token are mandatory
 
  \abstract
{The Einstein Telescope (ET), a wide-band, future third generation gravitational wave detector, is expected to have detection rates of $\sim 10^5 - 10^6$ binary black hole (BBH) detections and $\sim 7 \times 10^4$ binary neutron star (BNS) detections in one year. The coalescence of compact binaries with a total mass of 20 - 100 $M_{\odot}$, typical of BH-BH or BH-NS binaries, will be visible up to redshift $z\approx 20$ and even higher, thus facilitating the understanding of the dark era of the Universe preceding the birth of the first stars. The ET will therefore be a crucial instrument for population studies. We analysed the compact binaries originating in stars from (i) Population (Pop) I+II, (ii) Pop III, and (iii) globular clusters (GCs), with the single ET instrument, using the ET-D design sensitivity for the analysis. We estimated the constraints on the chirp mass, redshift, and merger rate as function of redshift for these classes of compact object binaries. We conclude that the ET as a single instrument is capable of detecting and distinguishing different compact binary populations separated in chirp mass - redshift space. While compact binaries originating in stars from Pop III are clearly distinguishable, owing to the separation in chirp mass - redshift space, the other two populations, Pop I+II, and GCs, can be distinguished with just 500 detections, corresponding to an observation time of $\sim 1$ hr.  The mass distribution characteristics of such different compact binary populations can also be estimated with the single ET instrument.} %Assuming that sufficient number of binaries will be detected from each of these populations, we also show  that such populations are distinguishable in the combined bulk detection. }
   \keywords{Gravitational waves; Stars: neutron, black holes; Methods: data analysis
             }

\maketitle
%
%-------------------------------------------------------------------

\section{Introduction}

The Einstein Telescope (ET) is planned to have a detection sensitivity down to 1Hz \citep{2008arXiv0810.0604H,2012CQGra..29l4006H}. It will thus have the ability to detect binary black holes (BBHs) of higher mass, between $10^2-10^4 M_{\odot}$ \citep{2011PhRvD..83d4020H,2011PhRvD..83d4021H,2011GReGr..43..485G,2010ApJ...722.1197A}. The ET will observe stellar mass binaries for a longer period of time in the detection band before their merger, due to improved low-frequency sensitivity. By the end of the most recent O3 run, LIGO-Virgo already has 90 gravitational wave events in the GWTC-3 catalogue \citep{2021arXiv211103606T}. Third generation detectors such as the ET  \citep{2011CQGra..28i4013H,2010CQGra..27s4002P}  or the Cosmic Explorer (CE) \citep{PhysRevD.91.082001,Abbott2017,2019BAAS...51g..35R} will have an even higher detection rate \citep{2020JCAP...03..050M}. The expected detection rates are  $\sim 10^5 - 10^6$ BBH detections and $\sim 7 \times 10^4$ binary neutron star (BNS) detections in one year \citep{2012PhRvD..86l2001R,2014PhRvD..89h4046R, 2019JCAP...08..015B}, based on the ET-D \citep{2011CQGra..28i4013H} design sensitivity. Assuming that the ET will have this design sensitivity, we investigate the prospects of using the ET as a single instrument, to analyse its capabilities for detecting and distinguishing different compact binary populations.

In an earlier work \citep{2021PhRvD.104d3014S} (SB1 hereafter),  we developed an algorithm to estimate the area of localisation and to provide constraints on chirp mass, redshift, and mass ratios by estimating their posterior distribution, for short duration gravitational wave signals from inspiraling compact binary systems. In a subsequent work \citep{2021arXiv210711198S} (SB2 hereafter), we took into account the effect of rotation of the Earth on the antenna pattern function to analyse long duration signals from coalescing low-mass compact binary systems.  We assumed the detector to be located at the Virgo site, and  we proceeded by analysing the signal every 5 minutes, estimating the constraints on the localisation and the binary parameters such as chirp mass, total mass, and redshift. In this analysis, as in the two previous works, we consider the ET as a single instrument rather than a part of network, to detect the gravitational radiation from an inspiraling compact binary system.

In the two preceding works, SB1 and SB2, we used a mock population, assuming that the mass distribution was the same for all distances. We also assumed the distributions of masses, distances, locations in the sky, and polarisations to be independent. We assumed a constant merger rate density with redshift, %uniform distribution in comoving volume 
and therefore did not take into account the dependence of the star formation rate (SFR) on redshift, or the delay between formation and coalescence.  Although such a population is not a realistic one, we chose such a simplified assumption in order to test the algorithm and to gauge the accuracy of the estimated parameters. We proceed to analyse more realistic populations of compact binary systems originating from Population (Pop) I and Pop II, Pop III,  and globular cluster (GC) populations in this work. 

In this analysis, we consider multiple populations of compact binary systems with masses varying over a wide range. The signal from low-mass binaries are expected to stay in the ET detection band for a long duration. We therefore use the algorithm developed in SB2 for analysing a  signal with such long duration  to estimate the parameters such as the chirp mass $\mathcal{M}$ and redshift $z$ of the compact binary systems. In \S \ref{ET} we describe the ET configuration, its antenna pattern function and the signal description. In \S \ref{plan_pop} we describe the methodology used for this analysis, while the details of the compact binary population models used to generate the mock population sets are given in Sect. \ref{pop_description}. In Sect. \ref{mock_pop_gen} we explain the parameters used to generate the mock population from the population models of compact binary systems originating from  I and  Pop II, Pop III,  and GC populations. We  present the results in 
\S \ref{results} and the conclusions in \S \ref{conc}.

\section{Detection of compact binary coalescence with the ET}\label{ET}

We consider the ET triangular configuration, consisting of three co-planar detectors, of equal arm lengths of 10 km,  with an opening angle of $60 ^{\circ}$. We assume the ET has the ET-D xylophone design \citep{2011CQGra..28i4013H}, which is a realistic version of ET-C \citep{2012CQGra..29l4006H} since it considers an improved noise model. Out of the multiple design configurations studied over time, ET-B \citep{2008arXiv0810.0604H} was the first basic design, which was based on a single cryogenic interferometer and covered full frequency range of interest. The design was next updated to the xylophone design, resulting in the ET-C sensitivity \citep{2012CQGra..29l4006H} in which each detector consisted of two interferometers, each with an opening angle of $60^{\circ}$. One interferometer was optimised for low frequencies while the other optimised for high frequencies. The single ET detector will thus have two interferometers, one for low frequencies and one for high frequencies. The final triangular design of the ET will thus have three such detectors, and six interferometers in total.

While observing a gravitational wave signal that stays in the detection band of the detector for a long duration, the change in the antenna response with the rotation of the Earth has to be taken into account. Following the detailed treatment of the antenna response given in \citep{PhysRevD.58.063001}, which takes into account the motion of the Earth, the time-dependent antenna response function for a single detector in the reference frame of the celestial sphere at time $t$ is given as:

\begin{subequations}\label{antenna_long}
\begin{equation}
    F_{+}(t) = \sin \eta \left[ a(t) \cos 2 \psi + b(t) \sin 2 \psi \right],
\end{equation}

\begin{equation}
    F_{\times}(t) = \sin \eta \left[ b(t) \cos 2 \psi - a(t) \sin 2 \psi \right],
\end{equation}
\end{subequations}
where,

\begin{subequations}
\begin{equation}
\begin{split}
    a(t)& = \frac{1}{16} \sin 2 \gamma (3 - \cos 2 \lambda)(3 - \cos 2 \delta)\cos[2(\alpha - \phi_r - \Omega_rt)]\\
    & - \frac{1}{4} \cos 2 \gamma \sin \lambda (3 - \cos 2 \delta) \sin[2(\alpha - \phi_r - \Omega_rt)]\\
    & + \frac{1}{4}\sin 2\gamma \sin 2 \lambda \sin 2 \delta \cos[\alpha - \phi_r - \Omega_rt]\\
    & - \frac{1}{2}\cos 2\gamma \cos \lambda \sin 2 \delta \sin[\alpha - \phi_r - \Omega_rt]\\
    & +\frac{3}{4} \sin 2 \gamma \cos^2 \lambda \cos^2 \delta,
\end{split}
\end{equation}
and

\begin{equation}
\begin{split}
    b(t) & = \cos 2 \gamma \sin \lambda \sin \delta \cos[2(\alpha - \phi_r - \Omega_rt)]\\
    & + \frac{1}{4} \sin 2 \gamma (3 - \cos 2 \lambda)\sin \delta \sin[2(\alpha - \phi_r - \Omega_rt)]\\
    & + \cos 2 \gamma \cos \lambda \cos \delta \cos[\alpha - \phi_r - \Omega_rt]\\
    & + \frac{1}{2} \sin 2 \gamma \sin 2 \lambda \cos \delta \sin[\alpha - \phi_r - \Omega_rt],
\end{split}
\end{equation}
\end{subequations}
where, $\alpha$ and $\delta $ are the right  ascension and the declination, respectively, of the gravitational wave source. $\psi$ is the polarisation angle, $\lambda$ is the latitude for the detector location, and $\Omega_r$ is Earth's rotational angular velocity. $\phi_r$ is the phase defining the position of the Earth in its diurnal motion at $t = 0$. The quantity $(\phi_r + \Omega_rt)$ is the local sidereal time at the detector site measured in radians. $\gamma$ determines the orientation of the detector arms and is measured counter-clockwise from the east to the bisector of the interferometer arms. $\eta$ is the angle between the interferometer arms. $\eta = 60^{\circ}$ for the ET in the triangular configuration.

Using Equation (\ref{antenna_long}), the antenna response functions can be calculated for any given instant of time $t$. The location of the ET detector for our analysis is chosen to be at the Virgo site.

According to general relativity, the two polarisations of the gravitational wave, produced by the gravitational wave signal from an inspiraling compact binary system, have a monotonically-increasing frequency and amplitude with the orbital motion radiating gravitational wave energy. The two polarisations for $t< t_c$  of the waveform for a binary merging at a distance $D_{L}$, with the chirp mass $\mathcal{M}$ described in Sect. 3 of \citep{Findchirp} are given as :

\begin{subequations}\label{hpluscross_ant}

\begin{equation}
\begin{split}
    h_{+}(t) = -\frac{1+\cos^{2}\iota}{2}\left(\frac{G \mathcal{M} }{c^2D_{L}}\right)\left(\frac{t_c-t}{5G\mathcal{M} /c^3}\right)^{-1/4}\\
    \times \cos\left[2\Phi_c + 2\Phi \left(t-t_c ; M,\mu\right)\right],
\end{split}
\end{equation}
\begin{equation}
\begin{split}
    h_{\times}(t) = -\cos\iota \left(\frac{G \mathcal{M}}{c^2D_{L}}\right)\left(\frac{t_c-t}{5G\mathcal{M}/c^3}\right)^{-1/4}\\
    \times \sin\left[2\Phi_c + 2\Phi \left(t-t_c ; M,\mu\right)\right],
\end{split}
\end{equation}
\end{subequations}
where, $c$ is the speed of light, $G$ is the gravitational constant. $\iota$ is the angle of inclination of the orbital plane of the binary system with respect to the observer. $\mu$ is the reduced mass of the binary system. The angle $\Phi \left(t-t_c ; M,\mu\right)$ is the orbital phase of the binary system. For a binary system composed of component masses $m_1$ and $m_2$, the chirp mass $\mathcal{M}$ is given as $\mathcal{M} = (m_1m_2)^{3/5}/M^{1/5}$, where $M = m_1+m_2$ is the total mass.  $t_{c}$ and $\Phi_{c}$ are the time and phase, respectively, of the termination of the waveform \citep{Findchirp}. The antenna response functions of one of the three detectors in ET, $F_+, F_\times$ are defined in Equation (\ref{antenna_long}). The strain in the detector $h(t)$ is then given as :

\begin{equation}\label{h_t_ant}
    h(t)= F_+ h_+(t+t_c-t_0) + F_{\times} h_\times(t+t_c-t_0),
\end{equation}
where $t_0$ is the time of coalescence in the detector frame and $(t_0 - t_c)$ is the travel of time from the source to the detector. The value of strain of the gravitational wave signal is obtained by substituting the values of the two polarisations from Equation (\ref{hpluscross_ant}) in Equation (\ref{h_t_ant}) :

\begin{equation}\label{ht}
\begin{split}
    h(t)= -\left(\frac{G\mathcal{M}}{c^2}\right)\left(\frac{\Theta}{4D_L}\right)\left(\frac{t_0-t}{5G\mathcal{M}/c^3}\right)^{-1/4}\\
    \times \cos\left[2\Phi_0 + 2\Phi \left(t-t_c ; M,\mu\right)\right]
\end{split},
\end{equation}
where, 
\begin{equation}\label{theta}
    \Theta\equiv 2 \left[F_{+}^{2}\left(1+\cos^{2}\iota\right)^{2} + 4F_{\times}^{2}\cos^{2}\iota\right]^{1/2},
\end{equation}
and,

\begin{equation}\label{snr_phase}
    2\Phi_0 = 2\Phi_c - \arctan\left(\frac{2F_\times\cos\iota}{F_+\left(1 + \cos^2\iota\right)}\right),
\end{equation}
with $0<\Theta<4$. The antenna response functions are slowly varying functions of time. Thus, in a given duration for which it can be assumed that the time of the signal in the detector bandwidth is short enough to ignore the change in the antenna response functions of the detector due to the rotation of the Earth,
\iffalse
, the Fourier transform of the gravitational wave signal amplitude  $h(t)$ in terms of frequency $f$ is \citep{PhysRevD.44.3819,TaylorGair2012,2010ApJ...716..615O}:

\begin{equation}
    |\tilde{h}(f)|= \frac{2c}{D_L}\left(\frac{5G\mu}{96c^3}\right)^{1/2}\left(\frac{GM}{\pi^2c^3}\right)^{1/3}\left(\frac{\Theta}{4}\right)f^{-7/6}\label{h_fourier}
\end{equation}
\fi
the signal to noise ratio (S/N) $\rho_j$ for $j = (1,2,3)$ for each of the three ET detectors, obtained using match filtering,  assuming that they have identical noise, is given as \citep{TaylorGair2012,LeeFinn96}:

\begin{equation}\label{snr}
\rho_j \approx 8 \Theta_j \frac{r_{0}}{D_{L}}
\left(\frac{\mathcal{M}_{z}}{\mathcal{M}_{BNS}}\right)^{5/6}\sqrt{\zeta\left(f_{max}\right),} 
\end{equation}
where  $\mathcal{M}_{z}= (1+z)\mathcal{M} $ is the redshifted chirp mass and $\mathcal{M}_{BNS}\approx 1.218 M_\odot $  is the chirp mass of an equal mass binary with each component mass being $1.4 M_{\odot}$.

\begin{equation}\label{zetafunc}
\zeta\left(f_{max}\right) = \frac{1}{x_{7/3}}\int^{2f_{max}}_{1}\frac{df \left( \pi M_{\odot}\right)^{2}}{\left(\pi f M_{\odot}\right)^{7/3}S_{h}\left(f\right)},
\end{equation}
where, $S_{h}\left(f\right)$ is the power spectral density (PSD) for the ET-D configuration for the ET-D noise curve \citep{2011CQGra..28i4013H} and,

\begin{equation}\label{x_7_3}
x_{7/3} = \int^{\infty}_{1}\frac{df \left( \pi M_{\odot}\right)^{2}}{\left(\pi f M_{\odot}\right)^{7/3}S_{h}\left(f\right)}.
\end{equation}
The characteristic distance sensitivity $r_0$ is:

\begin{equation}\label{detreach}
r^{2}_{0} = \frac{5}{192 \pi}\left(\frac{3 G}{20}\right)^{5/3}x_{7/3}\frac{M^{2}_{\odot}}{c^{3}},
\end{equation}
The frequency at the end of the inspiral phase $f_{max}$ is given as:

\begin{equation}\label{fmax}
f_{max} = 785\left(\frac{M_{BNS}}{M(1+z)}\right)  \;\rm Hz,
\end{equation}
where $M_{BNS}=2.8 M_\odot$ is the total mass of an equal mass binary with each component mass being $1.4 M_{\odot}$. We can define the combined effective S/N for the combined signal from three detectors as:

\begin{equation}\label{snreff}
\rho_{eff} = 8 \Theta_{eff} \frac{r_{0}}{D_{L}}\left(\frac{\mathcal{M}_{z}}{1.2 M_{\odot}}\right)^{5/6}\sqrt{\zeta\left(f_{max}\right),*}, 
\end{equation}
where the effective antenna response function  $\Theta_{eff}$ is:

\begin{equation}\label{thetaeff}
\Theta_{eff} = \left(\Theta_{1}^{2} + \Theta_{2}^{2} + \Theta_{3}^{2}\right)^{1/2}.
\end{equation}

\begin{figure}
\resizebox{\hsize}{!}{
\includegraphics[width=\columnwidth]{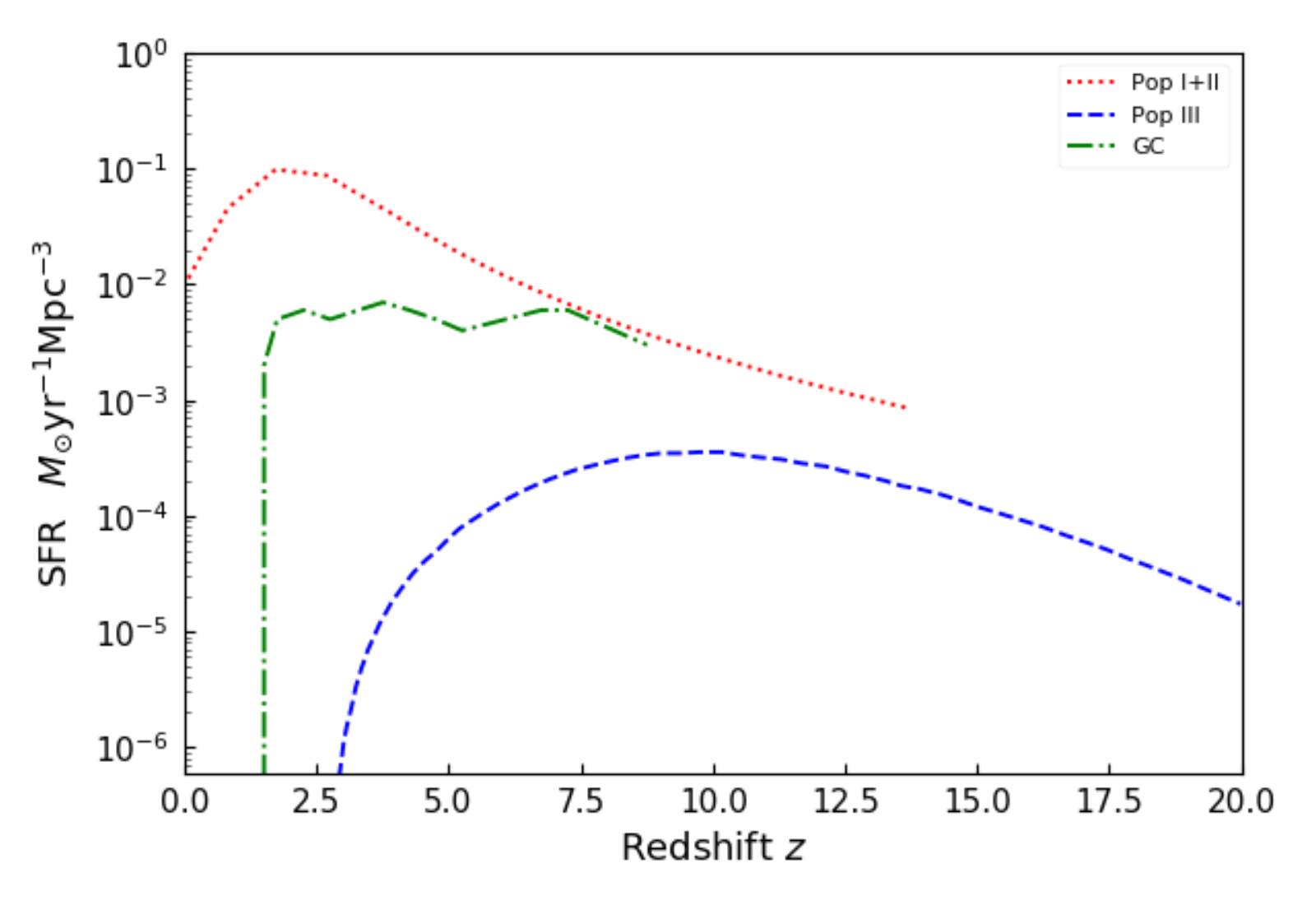}}
\caption{Star formation rate (SFR) used in the analysis for Pop I+II \citep{2017ApJ...840...39M}, Pop III \citep{2011A&A...533A..32D}, and the GC population \citep{2013MNRAS.432.3250K}. \label{fig:sfr_pop}}
\end{figure}

\begin{figure*}
\centering
\subfloat[\label{fig:chm_m30_cummu}]{\includegraphics[width=\columnwidth]{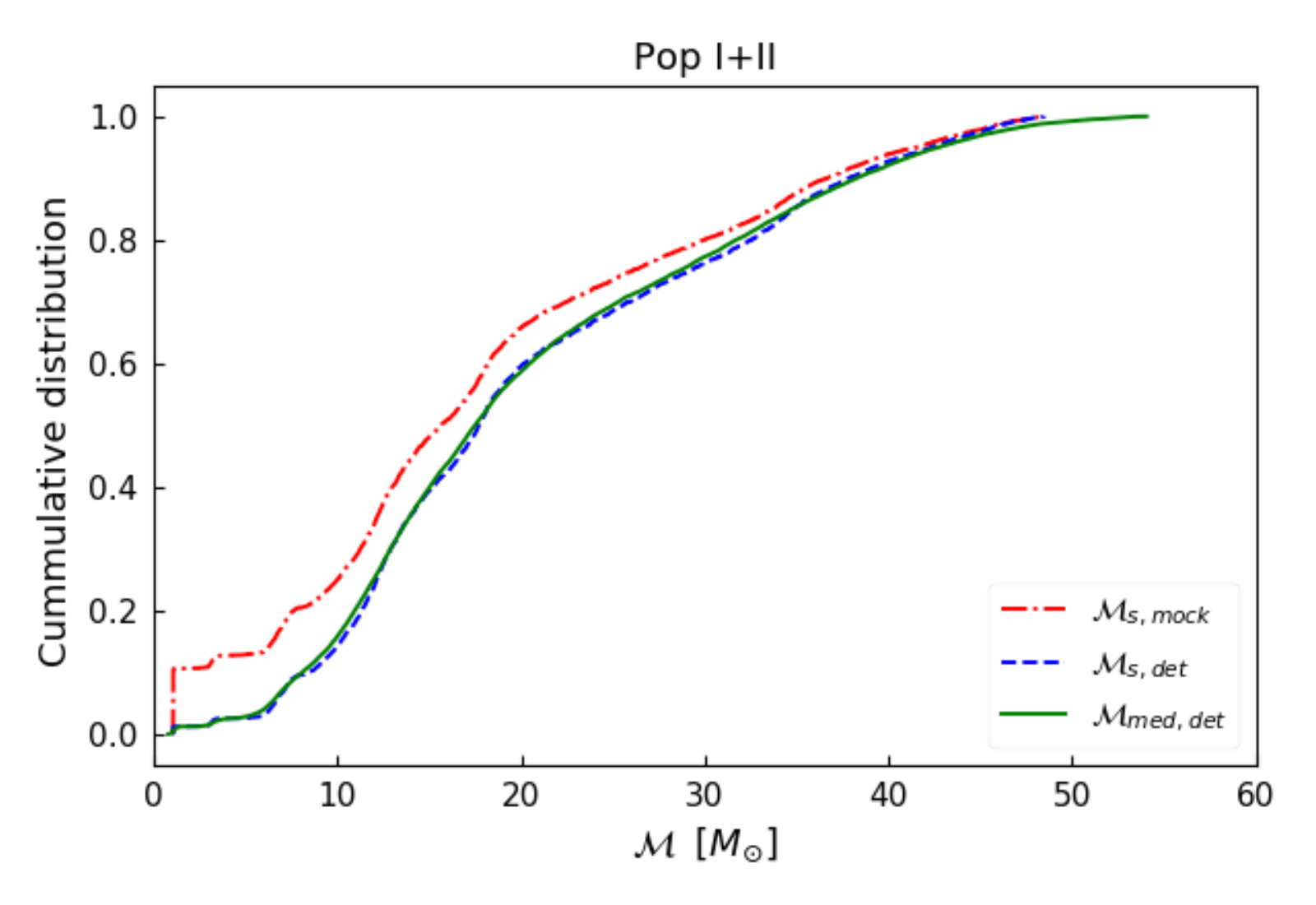}}\subfloat[\label{fig:red_m30_cummu}]{\includegraphics[width=\columnwidth]{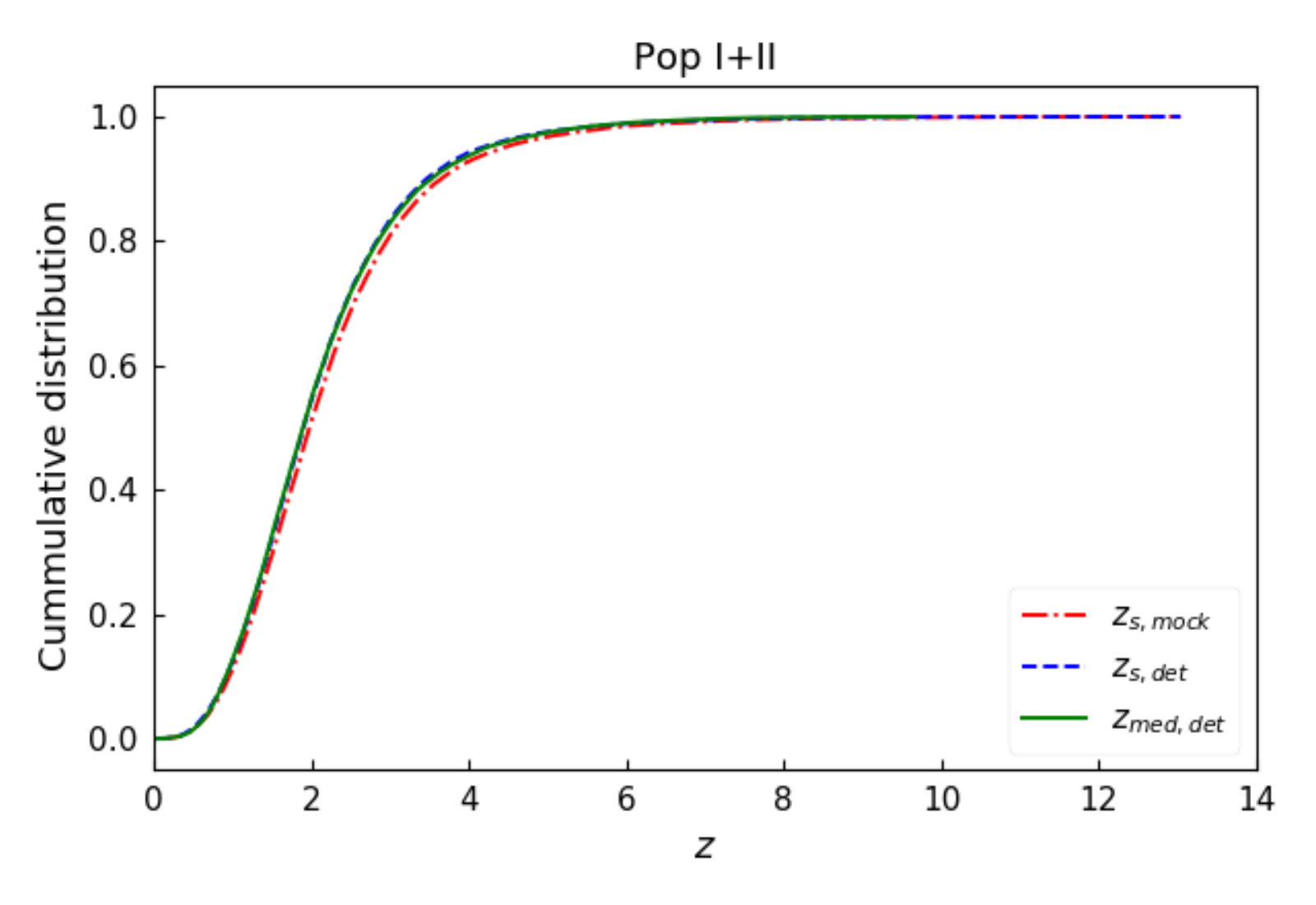}}\\
\subfloat[\label{fig:chm_pop3_cummu}]{\includegraphics[width=\columnwidth]{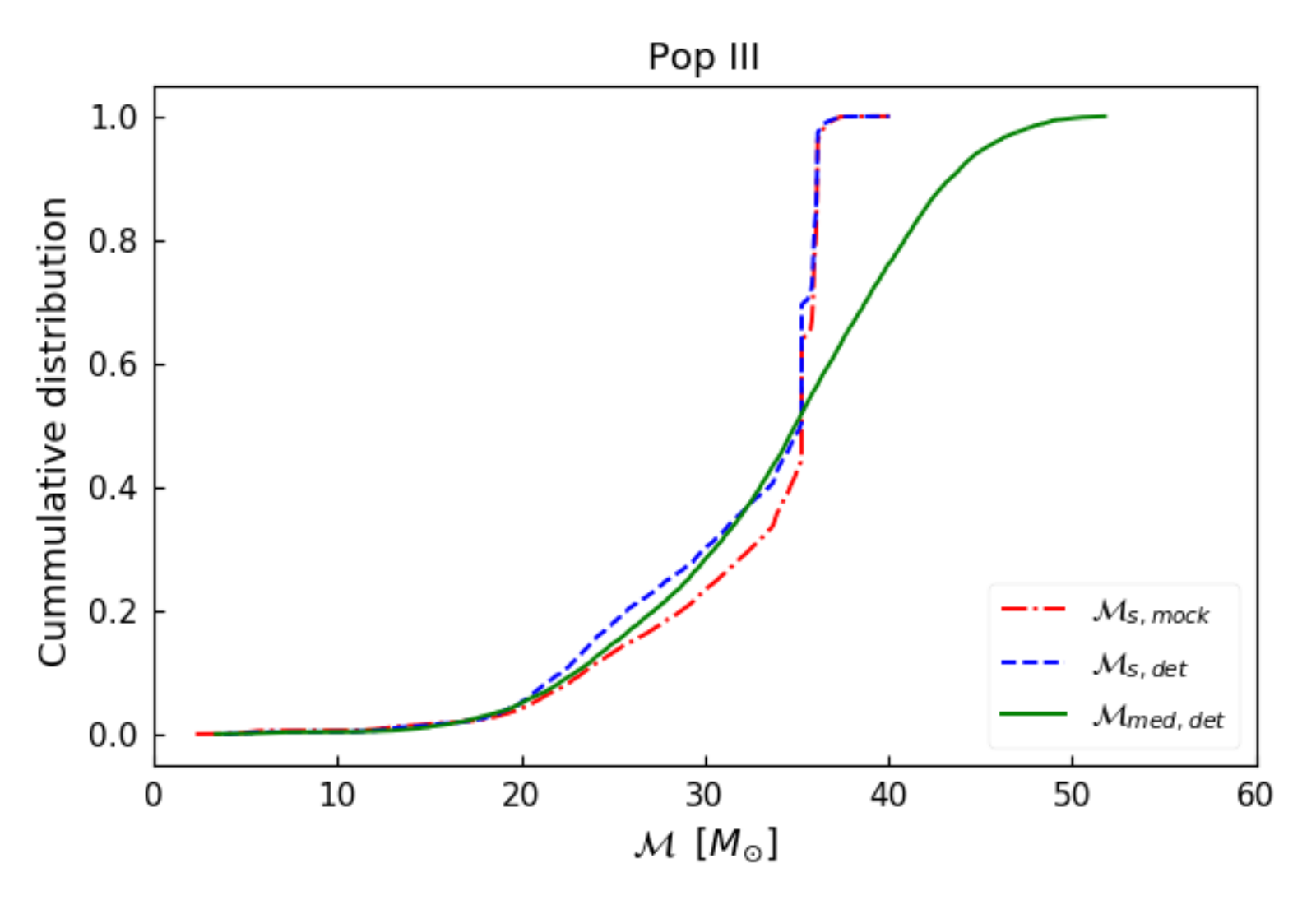}}\subfloat[\label{fig:red_pop3_cummu}]{\includegraphics[width=\columnwidth]{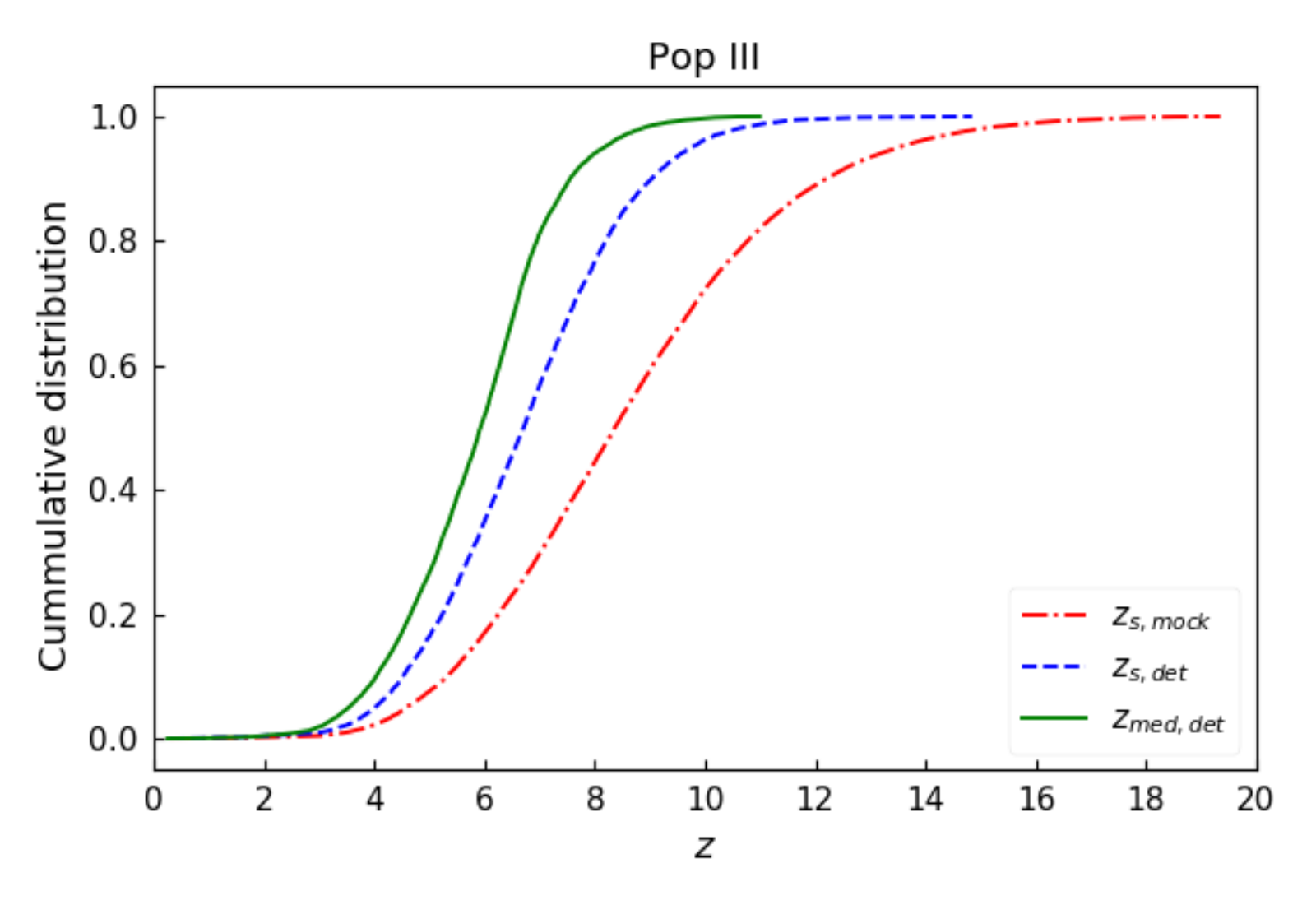}}\\
\subfloat[\label{fig:chm_cluster_cummu}]{\includegraphics[width=\columnwidth]{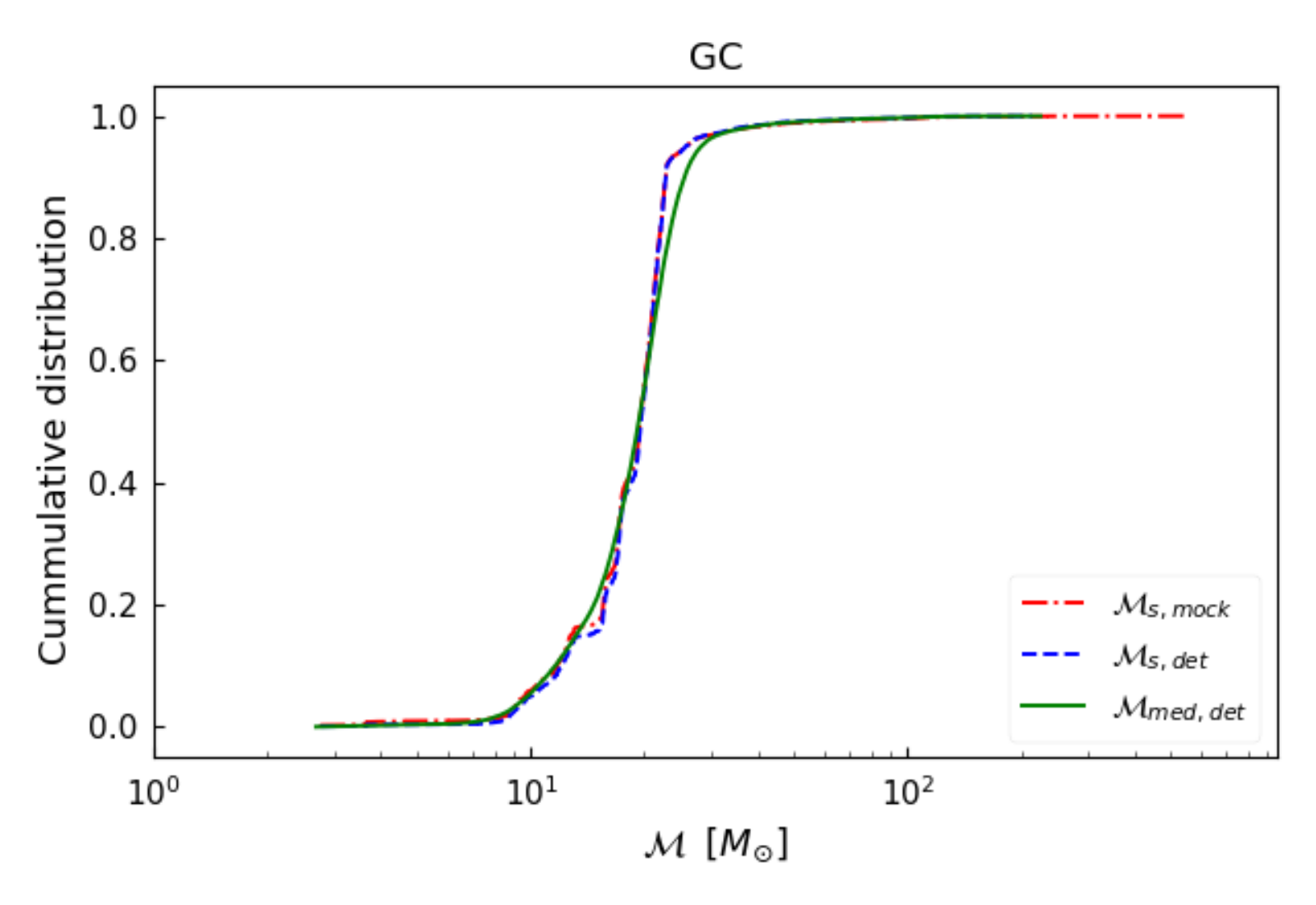}}\subfloat[\label{fig:red_cluster_cummu}]{\includegraphics[width=\columnwidth]{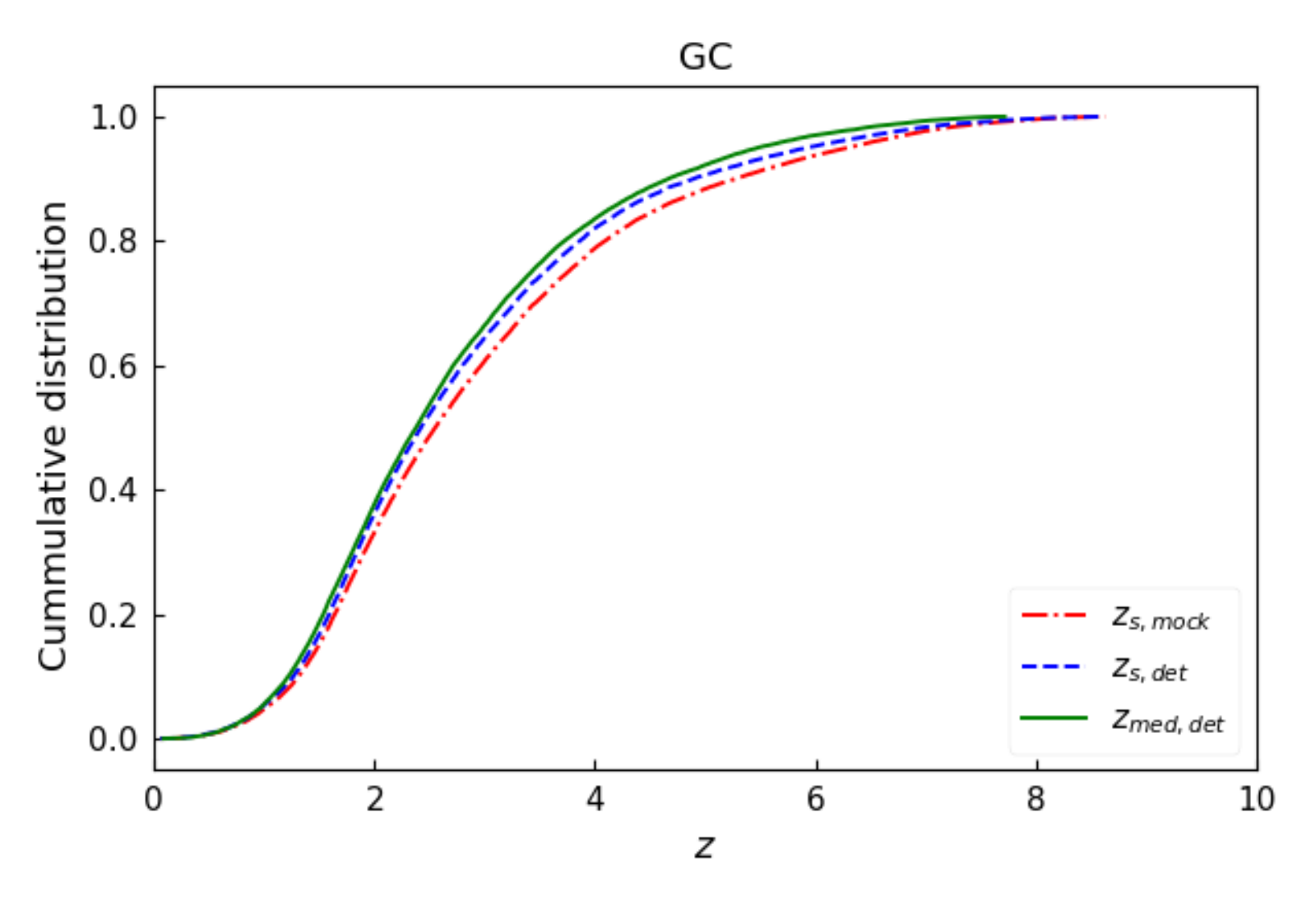}}\\
\caption{Cumulative distribution of the parameters: $\mathcal{M}$ (left) and $z$ (right) in the three populations. Top: Pop I+II. Middle: Pop III. Bottom: GC.}
\label{fig:cummulative_all}
\end{figure*}

\begin{figure*}
\centering
\subfloat[\label{fig:chm_m30}]{\includegraphics[width=0.9\columnwidth]{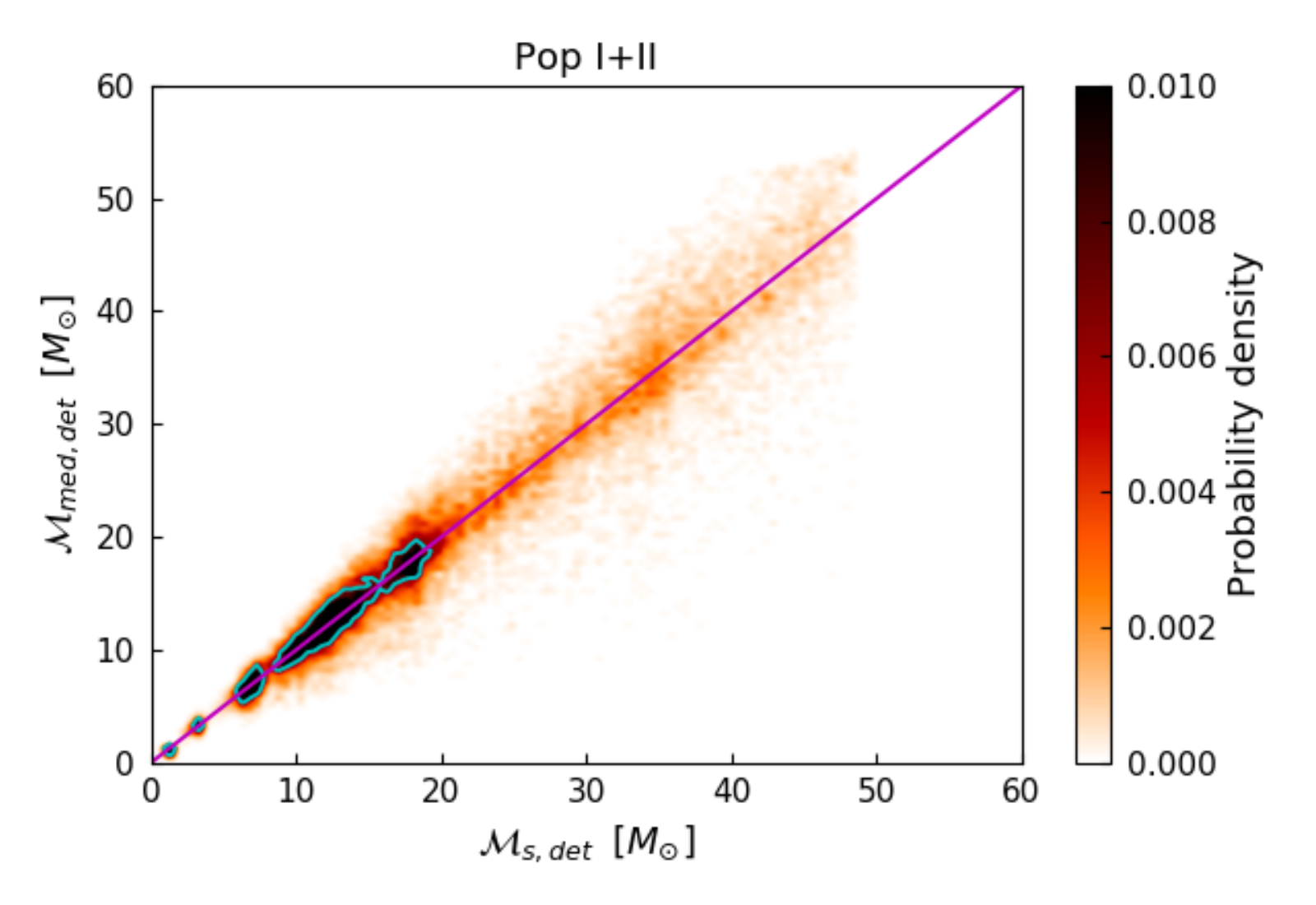}}\subfloat[\label{fig:red_m30}]{\includegraphics[width=0.9\columnwidth]{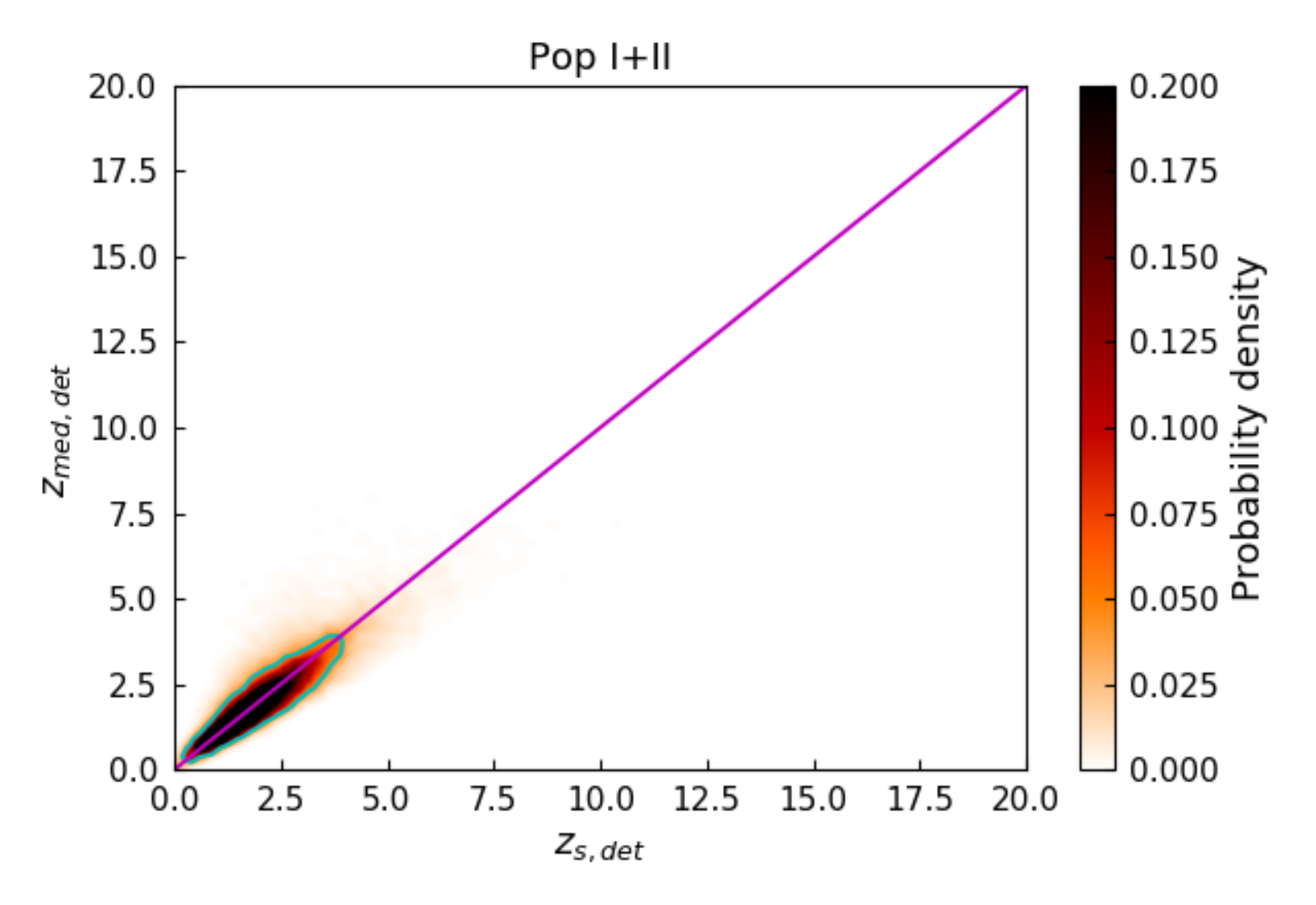}}\\
\subfloat[\label{fig:chm_pop3}]{\includegraphics[width=0.9\columnwidth]{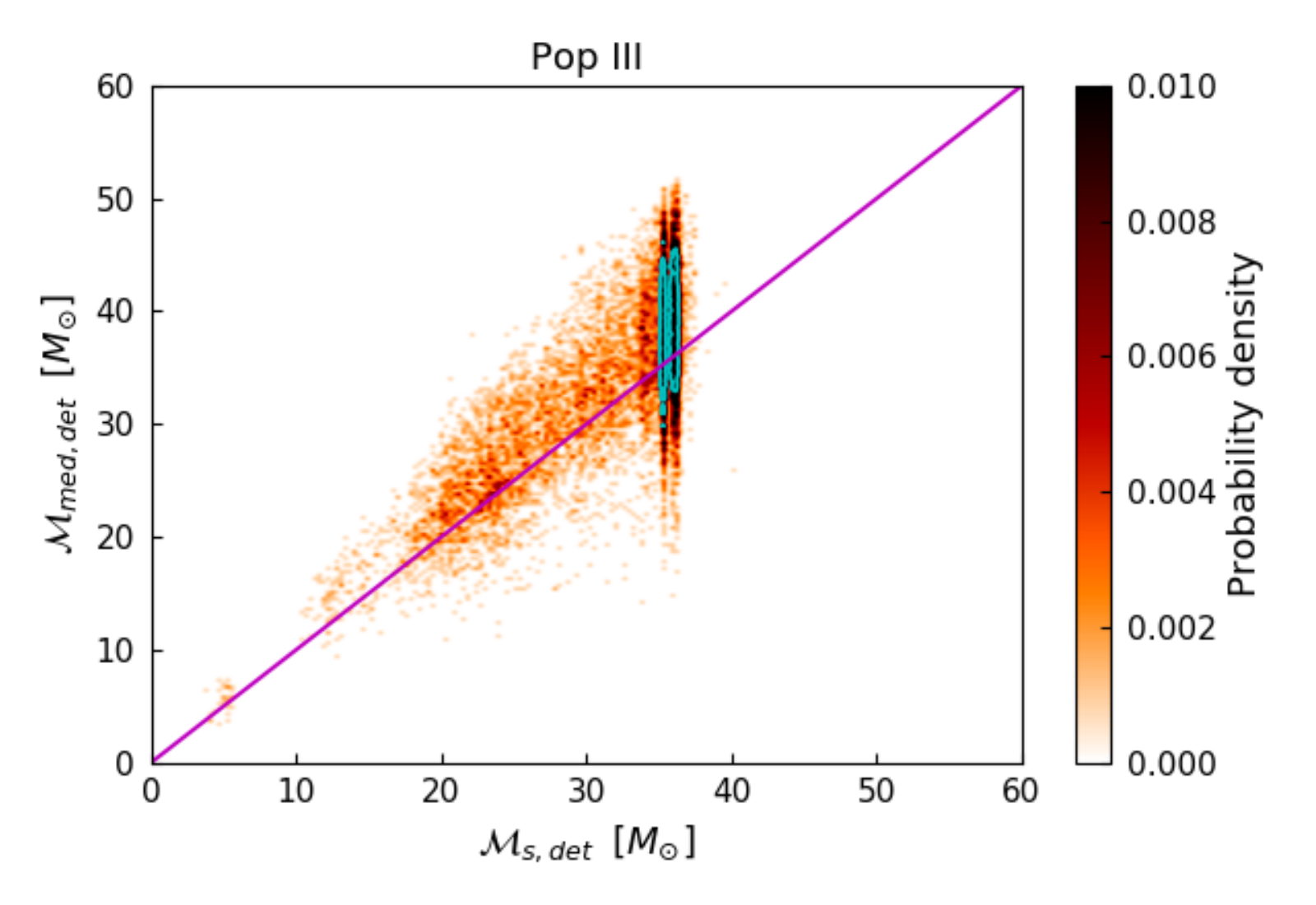}}
\subfloat[\label{fig:red_pop3}]{\includegraphics[width=0.9\columnwidth]{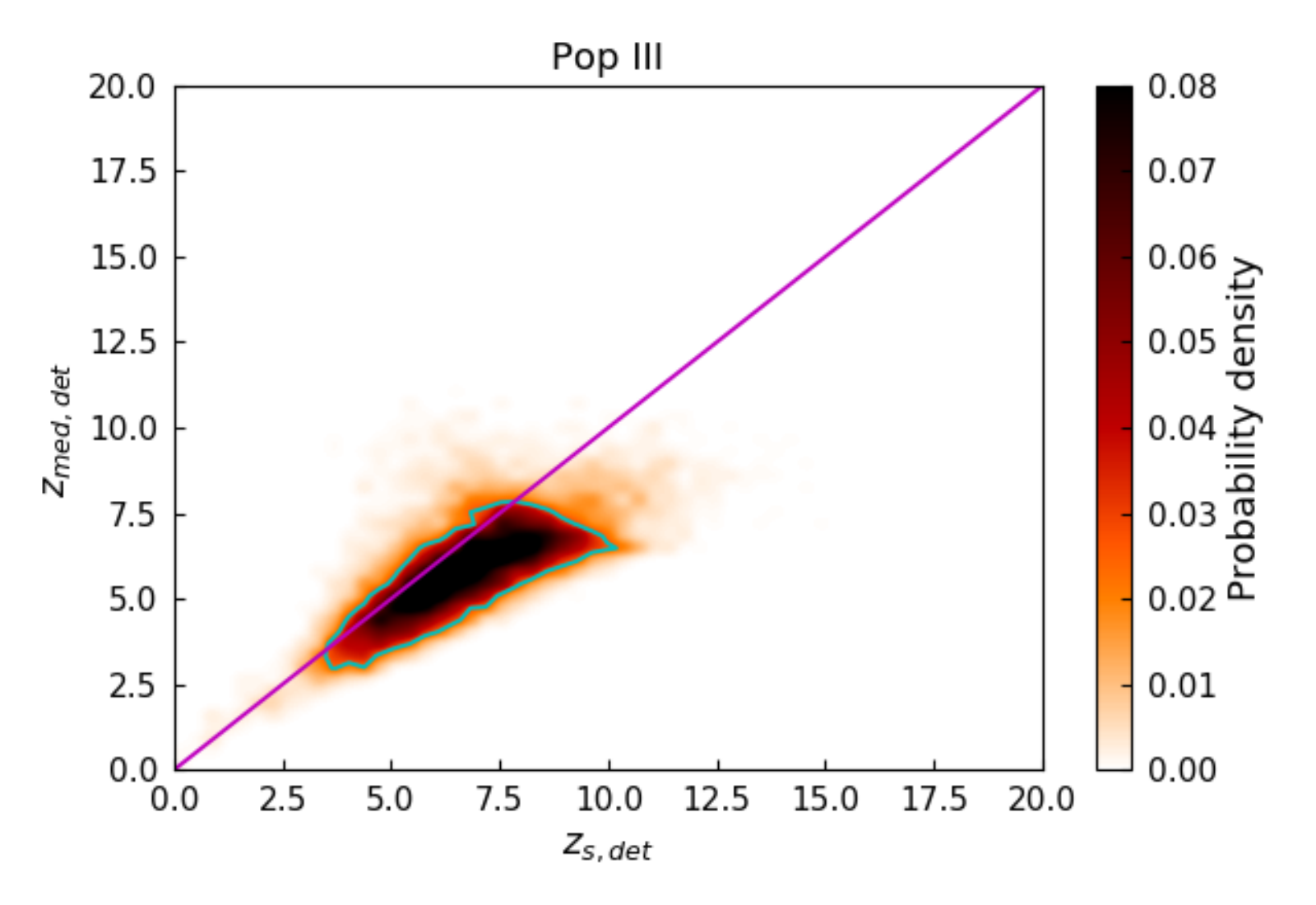}}\\
\subfloat[\label{fig:chm_cluster}]{\includegraphics[width=0.9\columnwidth]{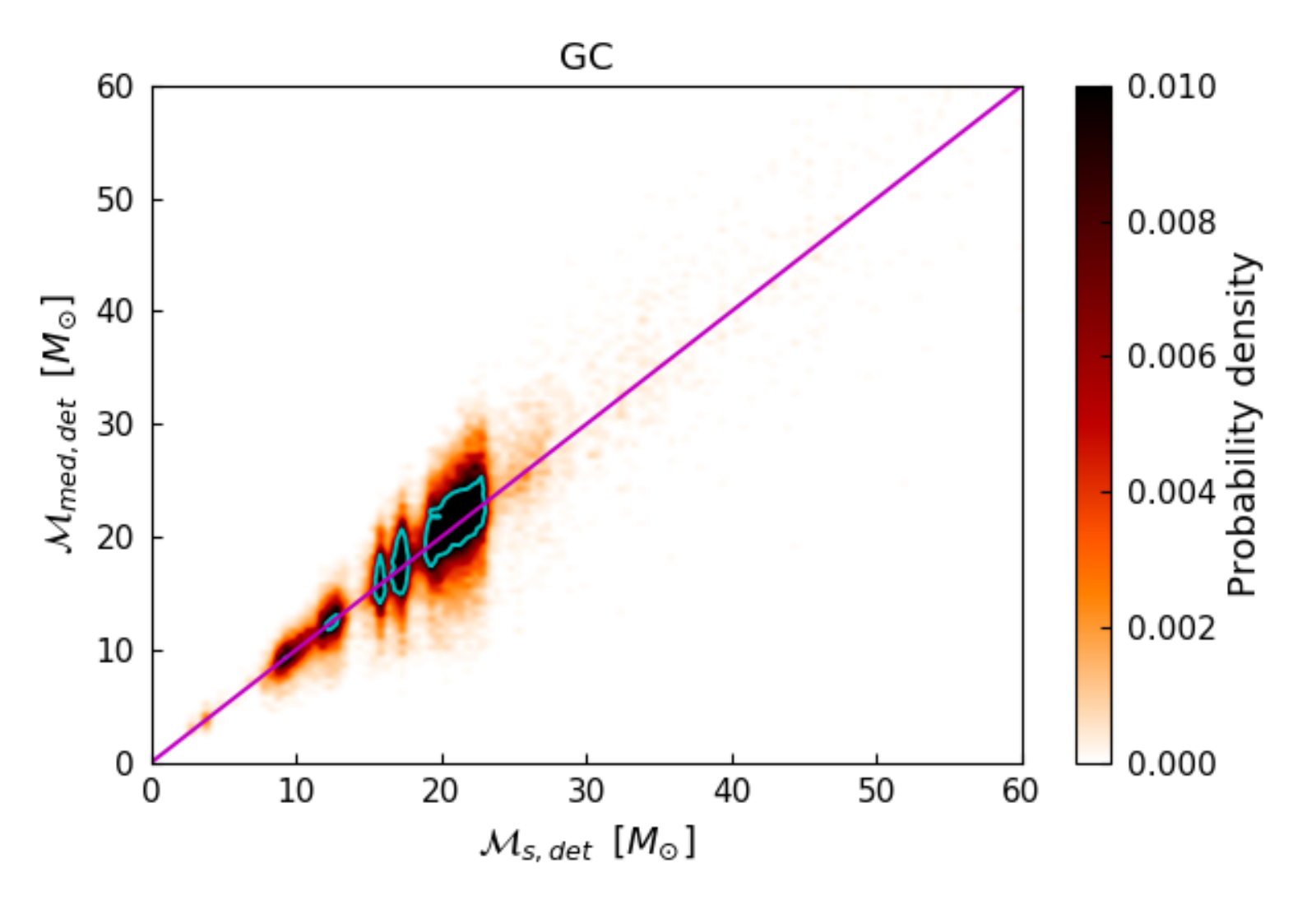}}\subfloat[\label{fig:red_cluster}]{\includegraphics[width=0.9\columnwidth]{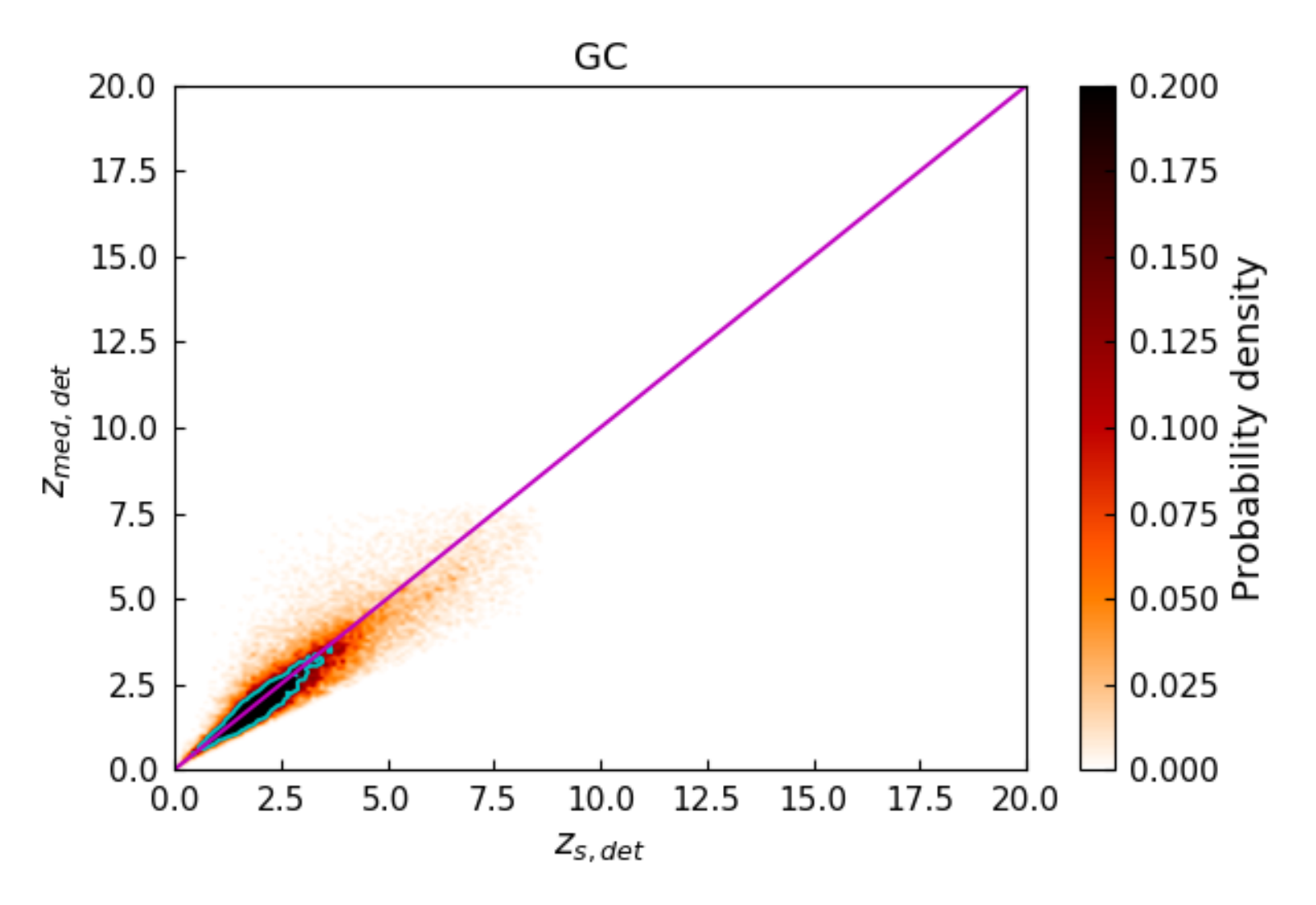}}\\
\caption{Density distribution of the estimated median values with respect to the actual values of the parameters: chirp mass $\mathcal{M}$ (left) and redshift $z$ (right) of the detected compact binary sources. Top: Pop I+II. Middle: Pop III. Bottom: GC. The magenta line is a reference for equal values of actual and estimated parameters. The blue contour encloses the region of 90\% probability.}
\label{fig:injvsmedian}
\end{figure*}

\section{Plan of the analysis}\label{plan_pop}

We use the algorithm developed for analysing long duration gravitational wave signals in SB2, to estimate the parameters such as chirp mass $\mathcal{M}$ and redshift $z$ of coalescing compact binary systems. We assume that in the case of detection of a coalescing binary system, the observables are as follows: (a) the three S/Ns $\rho^i_j$ defined by Equation (\ref{snr}) for each $i^{th}$ segment of the signal; (b) the phase of the strain $\Phi^i_{o,j}$ for $j = (1,2,3)$ corresponding to the three detectors defined in Equation (\ref{snr_phase}) for each $i^{th}$ segment of the signal. The quantity $\Phi_0$ is the best match phase obtained by maximising the matched filter output over the phase of the strain $h(t)$. The details are in \cite{Findchirp}; (c) the gravitational wave frequency at the start and end of each segment of the detected signal; (d) the redshifted chirp mass $\mathcal{M}_{z}$; and (e) the frequency at the end of the inspiral, corresponding to the innermost stable circular orbit, $f_{max}$. 

Given that the signal from low-mass binaries can stay in the ET detection band for a long duration, ranging from minutes to days, we take into account the change in the antenna pattern with the rotation of the Earth. Following the algorithm described in SB2, we use the response function \citep{PhysRevD.58.063001} for the celestial sphere frame of reference, given by Equation \ref{antenna_long}. Since antenna pattern is a slowly changing function of time, the change due to the rotation of Earth is negligible for a duration of 5 minutes. Thus, we choose to analyse the signal every 5 mins. While the lowest frequency for the  ET detection band is 1Hz, we fix the detection  at a threshold value of accumulated effective S/N $\rho_{eff}>8,$ and the S/N for $i^{th}$ segment in the $j^{th}$ detector $\rho^i_j> 3$ in at least one segment for $j = (1,2,3),$ corresponding to the three ET detectors comprising the single ET.  

The methodology of the analysis in this work is as follows: We consider compact binary systems from three sets of populations: (i) field binaries originating from Pop I and Pop II stars; (ii) binaries evolving in GCs; and (iii) binaries from Pop III stars. We construct a 'mock population' of compact binary systems merging within Hubble time from each of these three population sets. The chirp mass and redshift of these 'sources' are represented as $\mathcal{M}_{s,mock}$ and $z_{s, mock}$, respectively. We assume that the ET detector is located at the Virgo site and has the ET-D design sensitivity. For every compact binary system, from each of the three sets of mock population, the gravitational wave signal in the ET detection band is analysed by taking into account the change in the antenna pattern with the rotation of the Earth. The binary source is considered as 'detected' if it crosses a detection threshold set on the S/N. The chirp mass and redshift of these detected sources are represented as $\mathcal{M}_{s,det}$ and $z_{s, det}$, respectively. The probability distribution for chirp mass and redshift for each detected source is estimated using the algorithm described in SB2. (see Fig. 5 in SB2 for an example). The median values of these estimated distributions of chirp mass and redshift for each detected compact binary source are represented as $\mathcal{M}_{med,det}$ and $z_{med, det}$, respectively.

\section{Description of models used for generating mock populations}\label{pop_description}

To construct a mock population of compact binaries originating for Pop I+II stars, we use the model M30.B generated by \cite{2020A&A...636A.104B} using an upgraded version of population synthesis code StarTrack  \citep{2002ApJ...572..407B, 2008ApJS..174..223B}. For compact binaries from Pop III stars, we use the model FS1 generated by \citep{2017MNRAS.471.4702B}. The GC compact binaries population used in this analysis\footnote{with the permission of Prof. M. Giersz}, was generated by \citep{2017MNRAS.464L..36A}.

\begin{table*}
\centering
\caption{Details of three population sets simulated for the analysis.\label{tab:pop_details}}
\begin{tabular}{c|cccccc}
\hline
Type of population & $N_{yr}$ & $N_{mock}$ & $T_{mock}$ & $N_{det}$ & $(z_{min}, z_{max})_{mock}$ & $(\mathcal{M}_{min}, \mathcal{M}_{max})_{mock}$\\
 & & & [yr] & & &[$M_{\odot}$]\\
\hline
Pop I+II & 4092783 & 25862 & 0.006 & 20528 & (0.03, 13.04) & (0.96, 48.47)\\
Pop III & 2429 & 23990 & 9.878 & 9683 & (0.28, 19.36) & (2.27, 40.04)\\
GC & 362810 & 26461 & 0.073 & 22937 & (0.06, 8.63) &  (2.77, 554.63)\\
\hline
%\hline
\end{tabular}
\tablefoot{$N_{yr}$ is the expected number of mergers per year, $N_{mock}$ is the simulated number of binaries, $T_{mock}$ is the time during which the binaries in the mock population are expected to merger, and $N_{det}$ is the number of binaries that are detected, based on the chosen detection threshold. }
\end{table*}

\subsection{Pop I+II compact binaries}\label{subsec:pop1pop2}

\cite{2020A&A...636A.104B} generated a broad set of models of compact binary systems with an upgraded version of population synthesis code StarTrack  \citep{2002ApJ...572..407B, 2008ApJS..174..223B}. We use the model M30.B to generate our population of compact binaries originating from Pop I+II stars. 
The characteristic features of this model are described in detail in \citep{2020A&A...636A.104B} and can we summarise them as follows: (i) Supernova engine model for NS and BH mass: This model uses a rapid supernova engine model as mentioned in \citep{2012ApJ...749...91F}. It creates a mass gap in the range $(2-5) M_{\odot}$ between NSs and BHs. For this model, the minimum BH mass is  assumed to be $2.5 M_{\odot}$; (ii) Pair-instability  supernovae (PSNs) and pair-instability pulsation supernovae (PPSNs): this model allows PSNs  for stars  with  He  core  mass  in  the range $(65-135) M_{\odot}$, thus entirely disrupting  massive  stars, while for the low-mass stars with He core mass in the range $(40-65) M_{\odot}$, it allows weak  PPSNs, resulting in only up to 50\% of mass loss  calculated  by \citep{2019ApJ...887...72L}, so that in this model, the  maximum mass of a post PPSN\ star may reach $55.6 M_{\odot}$; (iii) Neutrino mass loss: this model allows 1\% neutrino mass loss to all BHs, while it allows 10\% neutrino mass loss at NS formation; (iv) Natal kicks: this model assigns low natal BH kicks for low-mass BHs and no natal kicks for high-mass BHs. High kicks are assigned to NSs; (v) Accretion: in this model it is assumed that 50\% of mass lost by the donor during Roche lobe overflow (RLOF) is accreted on to the companion star if the companion star in not a compact object. Accretion on compact objects (NSs or BHs) is allowed to be at a super-Eddington rate, as presented by \citep{2020MNRAS.491.2747M} A 5\% rate of Bondi-Hoyle accretion is allowed onto NSs and BHs during CE evolution; and (vi) Winds: the model uses the massive star winds \citep{2001A&A...369..574V} in addition to luminous blue variable (LBV) winds \citep{2010ApJ...714.1217B}. It is assumed that LBV mass loss is independent of metallicity. In the case of Wolf–Rayet  (W-R) stars, the model assumes a prescription that is a combination of the \cite{1998A&A...335.1003H} wind rate estimate, which takes into account W-R wind clumping, and \cite{2005A&A...442..587V} wind metallicity dependence.

This model is obtained from the StarTrack site\footnote{http://www.syntheticuniverse.org/} and contains result from 32 simulations of binary evolution with various metallicities $Z_i$, where $i=(1:32)$. The metallicity grid values used in the model are :

\begin{equation}\label{Zi_m30}
\begin{split}
        Z =  \ \ & 0.0001, 0.0002, 0.0003, 0.0004, 0.0005,  0.0006, \\ 
        & 0.0007, 0.0008, 0.0009,  0.001,  0.0015,  0.002, 0.0025, \\ 
        &  0.003, 0.0035, 0.004, 0.0045, 0.005, 0.0055, 0.006, \\ 
        & 0.0065, 0.007, 0.0075, 0.008, 0.0085, 0.009, 0.0095, \\
        &  0.01, 0.015, 0.02, 0.025,0.03.
\end{split}
\end{equation}
Each simulation is performed with the same initial conditions and is a result of simulating the same number (2 million) of massive binaries.

\subsection{Pop III compact binaries}

\cite{2017MNRAS.471.4702B} generated two models for the population of compact binary systems from the first, metal-free  Pop III stars in the Universe using the population synthesis code StarTrack \citep{2002ApJ...572..407B, 2008ApJS..174..223B}. They approximated the evolution of these Pop III stars with their metallicity model of

\begin{equation}\label{Zi_pop3}
    Z = 0.0001,
\end{equation}
assuming no mass loss via stellar winds, since mass loss in stellar wind is expected to be negligible for massive Pop III stars \citep{2001ApJ...550..890B}. The radial expansion of the $Z = 0.0001$ metallicity model is constrained to match the upper limit of expansion of $Z = 0$ stars using the evolutionary model for Pop III stars calculated by \citep{2001A&A...371..152M}.

The initial conditions of the Pop III stars were defined based on the models obtained by \citep{2016MNRAS.456..223R} using N-body simulation. Two models were chosen representing two different sizes of a gas cloud (mini-halo). These two models considered the formation of Pop III  stars in mini haloes of sizes $\sim 2000$ AU and $10-20$ AU. For our analysis we use the results from the model FS1 \citep{2017MNRAS.471.4702B}, which assumes the formation of Pop III stars from a mini halo of size $\sim 2000$ AU \citep{2013MNRAS.433.1094S}. Following the parameters specified by \citep{2013MNRAS.433.1094S}, the number density of the gas medium is chosen to be $10^6\rm{cm}^{-3}$. The resulting population of metal-free binaries (initial mass function, mass ratio, orbital separations, and eccentricities) are presented in detail in \cite{2017MNRAS.471.4702B}. The population was taken from the StarTrack\footnote{http://www.syntheticuniverse.org/} website, which hosts the results of StarTrack simulations.

\subsection{Globular cluster population}

For data pertaining to merging BBHs produced by GCs, we made use of the results from \textsc{MOCCA}-Survey Database I \citep{2017MNRAS.464L..36A}. This database comprises about 2000 simulated GC models that are characterised by different initial parameters. The long-term evolution of these cluster models was carried out using the \textsc{MOCCA} code \citep{hypki2013,2013MNRAS.431.2184G}. To compute the dynamical evolution of a spherically symmetrical star cluster, the \textsc{MOCCA} code uses the orbit-averaged Monte Carlo treatment for relaxation \citep{henon1971b,stod1986,giersz2001}. To determine the outcome of binary-single and binary-binary encounters, it uses the \textsc{FEWBODY} code \citep{fregeau2004}, which is a direct N-body integrator for small-N encounters. For stellar and binary evolution of stars, the version of MOCCA used in \citep{2017MNRAS.464L..36A} utilised the prescriptions proved by the (\textsc{SSE}) and binary (\textsc{BSE}) codes \citep{hurley2000,hurley2002}. \textsc{MOCCA} also uses realistic treatment for the escape process in tidally limited star clusters as described by \citep{fukushige2000}.

The 2000 star cluster models in \textsc{MOCCA}-Survey Database I had a different initial number of objects, metallicities, binary fractions, central densities, half-masses, and tidal radii (see Table 1 in \citep{2017MNRAS.464L..36A}). The initial stellar masses in these star cluster models are sampled from the initial mass function given by \citep{2001MNRAS.322..231K}, and have values between  $0.08 \ M_{\odot}$ and$100 \ M_{\odot}$. For each initial model in \textsc{MOCCA}-Survey Database I, two models were simulated with different prescriptions for BH natal kicks. In one case, the BH masses and natal kicks are computed using the mass fallback prescription from \citep{2002ApJ...572..407B}. In the other case, the BHs are given the same natal kicks as NSs, which follow a Maxwellian distribution with $\sigma = 265\,\text{km}\text{s}^{-1}$ \citep{hobbs2005}. The simulations assumed the metallicity values to be: \begin{equation}\label{Zi_cluster}
    Z = 0.0002, 0.001, 0.005, 0.006, 0.02,
\end{equation}
for which 63, 831, 487, 64, and 503 models were simulated, respectively. The data for merging BBHs were taken from about 1000 of these models in which the mass fallback prescription from \citep{2002ApJ...572..407B} was used to determine BH masses and natal kicks. The present-day observable properties of these stellar cluster models are broadly consistent with the observed values for Galactic and extra-galactic GCs \citep{askar2018,leveque2021}.

\cite{2017MNRAS.464L..36A} found 15134 coalescing BBHs that escape the clusters and 3000 BBHs that merged inside the cluster. The lower limit for the local merger rate of BBHs computed from these GC simulations is $5.4$ ${\rm Gpc}^{-3}\,{\rm yr}^{-1}$. The cluster models simulated by \citep{2017MNRAS.464L..36A} did not have the latest prescriptions for stellar evolution of BH progenitors connected with wind mass loss, PSNs and PPSNs, and the supernova engine model for NS and BH mass (see Sect. \ref{subsec:pop1pop2}). For low metallicities, the maximum BH mass formed through single stellar evolution in \citep{2017MNRAS.464L..36A} is about 30 $M_{\odot}$.
Additionally, these simulations do not take into account post-Newtonian corrections during few-body encounters involving BHs. Such corrections can lead to gravitational wave capture or the merger of BHs \citep{samsing2018a,rodriguez2018b,samsing2018b} inside star clusters. Despite these shortcomings, the merger rates of merging BBHs from \citep{2017MNRAS.464L..36A} are consistent with lower limits from newer GC simulations in which these updates have been implemented \citep{kremer2020,2022PhRvD.105b3004B,2022MNRAS.511.5797M}.

\subsection{Cosmological model}

We assume flat cosmology, as assumed in the preceding works SB1 and SB2,  with $\Omega_m=0.3$,  $\Omega_m + \Omega_\Lambda = 1, $ $\Omega_k = 0$, $H_0 = 67.3 \; \rm km s^{-1} \;\rm Mpc^{-1}$ \citep{2015PhRvD..92l3516A}, and the relation between the luminosity distance $D_L$ and redshift $z$ is obtained using the analytic approximation given by \citep{Adachi2012}. For this cosmology, the variation of comoving volume $V_c$ with redshift $z$ is given as 

\begin{subequations}
\begin{equation}\label{dVdz_pop}
  \frac{dV}{dz} = 4\pi D_H\frac{D_L^2}{(1+z)^2E_z},
\end{equation}
where
\begin{equation}
D_H = c/H_0 \;  {\rm and}  \; E_z = \sqrt{\Omega_m(1+z)^3 + (1-\Omega_m)}.
\end{equation}
\end{subequations}

\begin{figure*}
     \centering
     \includegraphics[scale = 0.5]{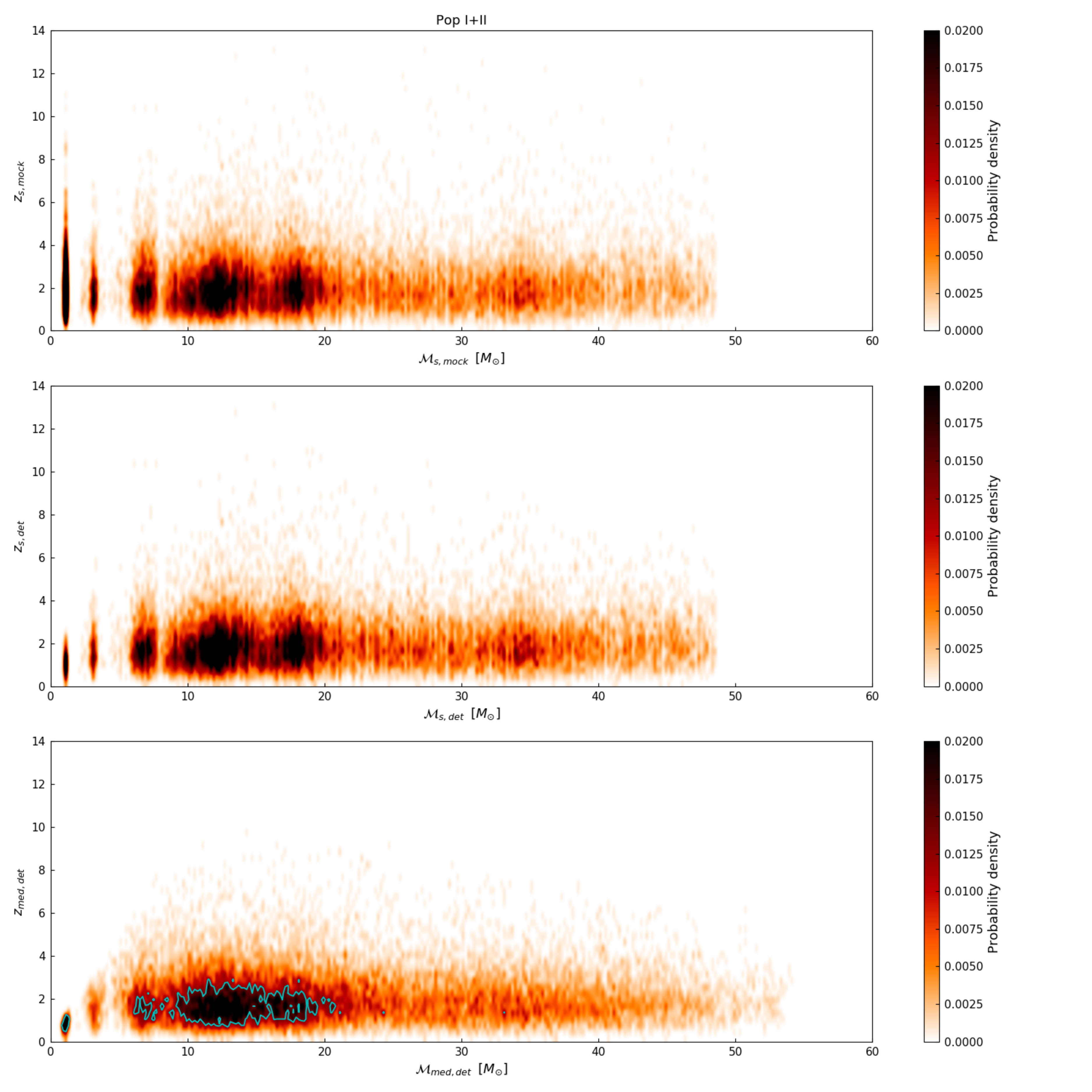}
\caption{Probability density of chirp mass and redshift in the $(\mathcal{M}-z)$ space for the three sets of parameters for Pop I+II binaries. Top: Actual parameters of the sources in the mock population, $\mathcal{M}_{s, mock}$ and $z_{s, mock}$. Middle: Actual parameters of the binaries that were detected, $\mathcal{M}_{s, det}$ and $z_{s, det}$. Bottom: Estimated medians of the parameters for each of the detected compact binaries  $\mathcal{M}_{med, det}$ and $z_{med, det}$. The blue contour encloses the region of 90\% probability.}
\label{fig:chm_z_m30}
\end{figure*}

\begin{figure*}
     \centering
     \includegraphics[scale = 0.5]{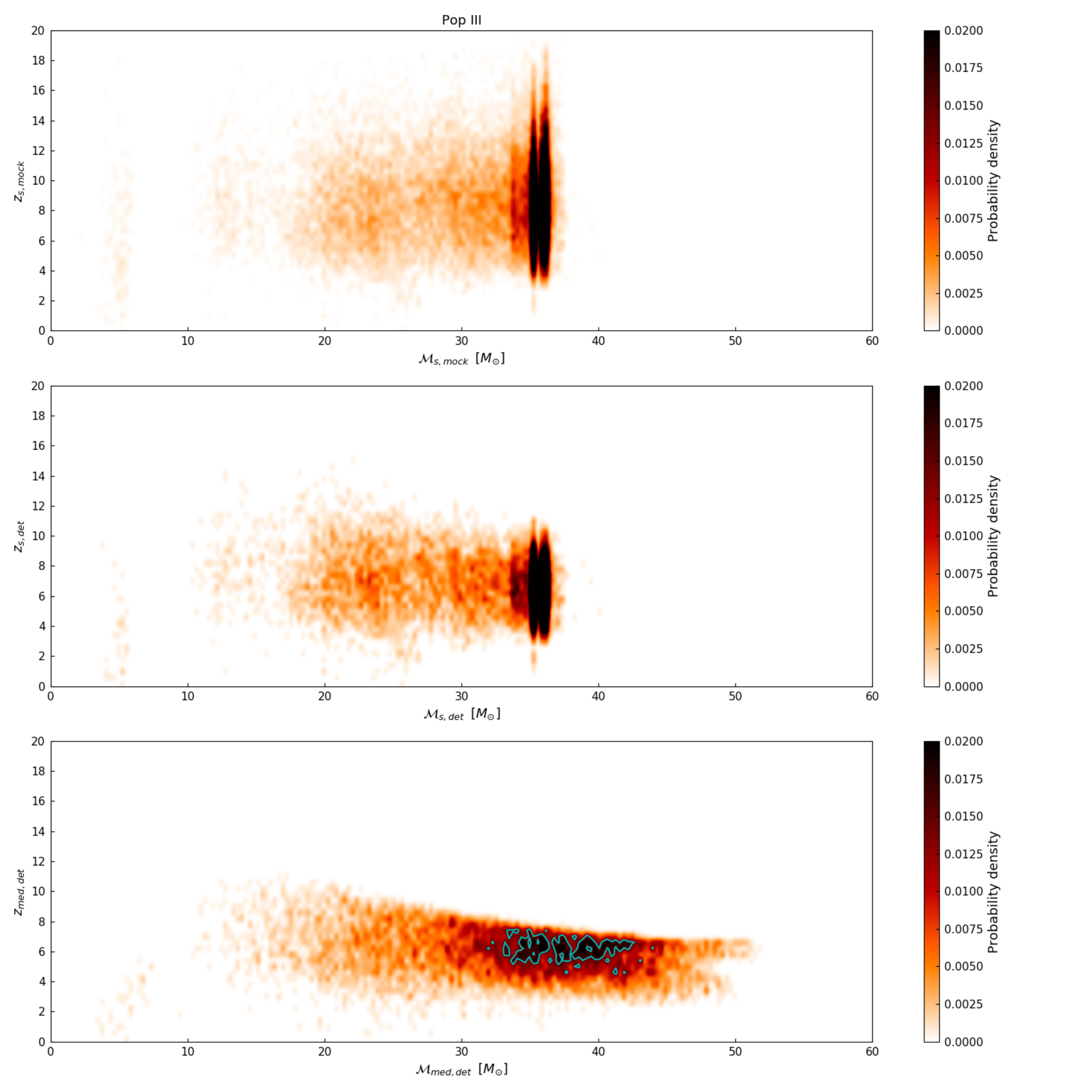}
\caption{Probability density of chirp mass and redshift in the $(\mathcal{M}-z)$ space for the three sets of parameters for Pop III binaries. Top: Actual parameters of the sources in the mock population, $\mathcal{M}_{s, mock}$ and $z_{s, mock}$. Middle: Actual parameters of the binaries that were detected, $\mathcal{M}_{s, det}$ and $z_{s, det}$. Bottom: Estimated medians of the parameters for each of the detected compact binaries,  $\mathcal{M}_{med, det}$ and $z_{med, det}$. The blue contour encloses the region of 90\% probability.}
\label{fig:chm_z_pop3}
\end{figure*}

\begin{figure*}
     \centering
     \includegraphics[scale = 0.5]{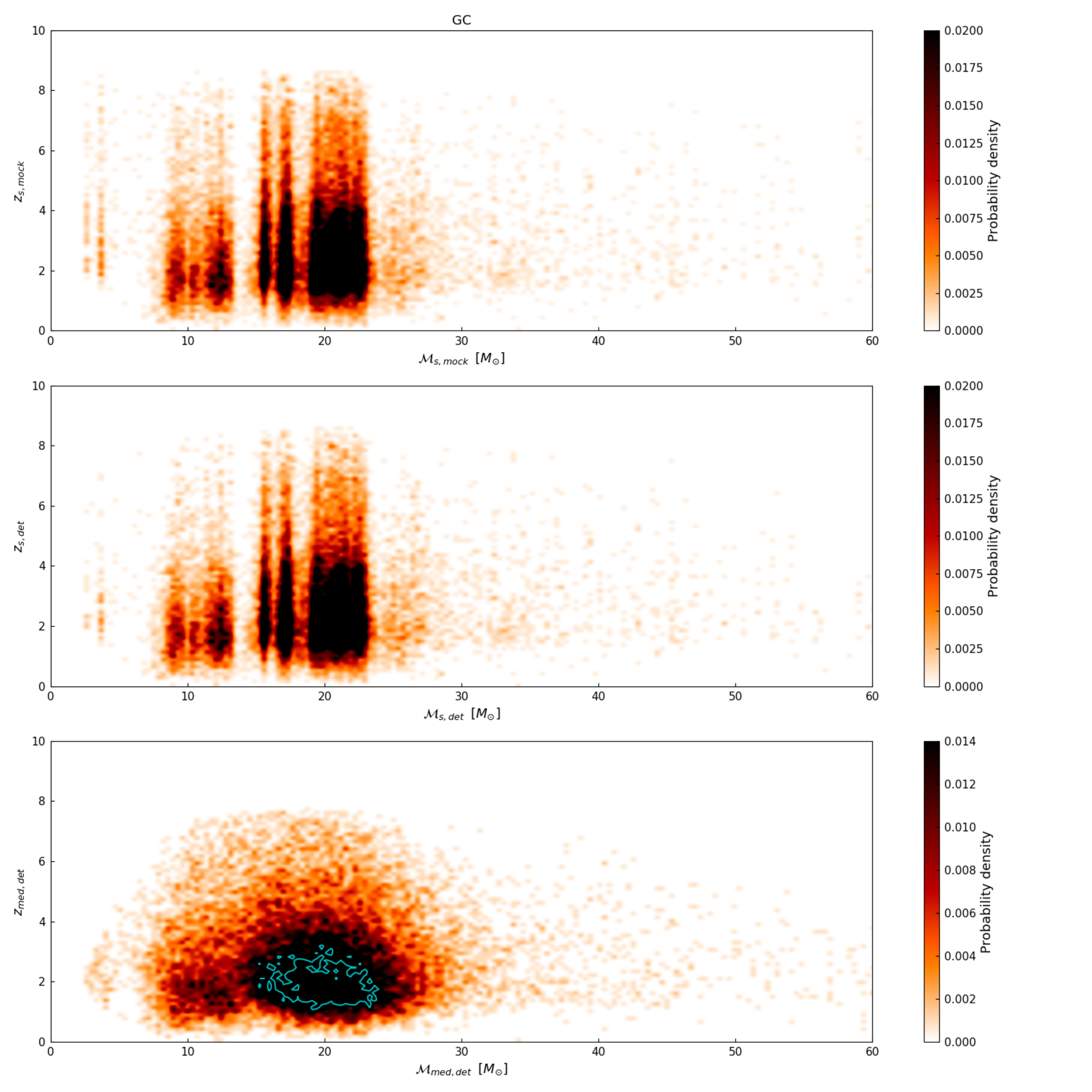}
\caption{Probability density of chirp mass and redshift in the $(\mathcal{M}-z)$ space for the three sets of parameters for GC binaries. Top: Actual parameters of the sources in the mock population, $\mathcal{M}_{s, mock}$ and $z_{s, mock}$. Middle: Actual parameters of the binaries that were detected $\mathcal{M}_{s, det}$ and $z_{s, det}$. Bottom: Estimated medians of the parameters for each of the detected compact binaries, $\mathcal{M}_{med, det}$ and $z_{med, det}$. The blue contour encloses the region of 90\% probability.}
\label{fig:chm_z_cluster}
\end{figure*}

\section{Mock source catalogue for compact binaries}\label{mock_pop_gen}

\begin{figure*}
\subfloat[\label{fig:rate_m30}]{\includegraphics[width=0.9\columnwidth]{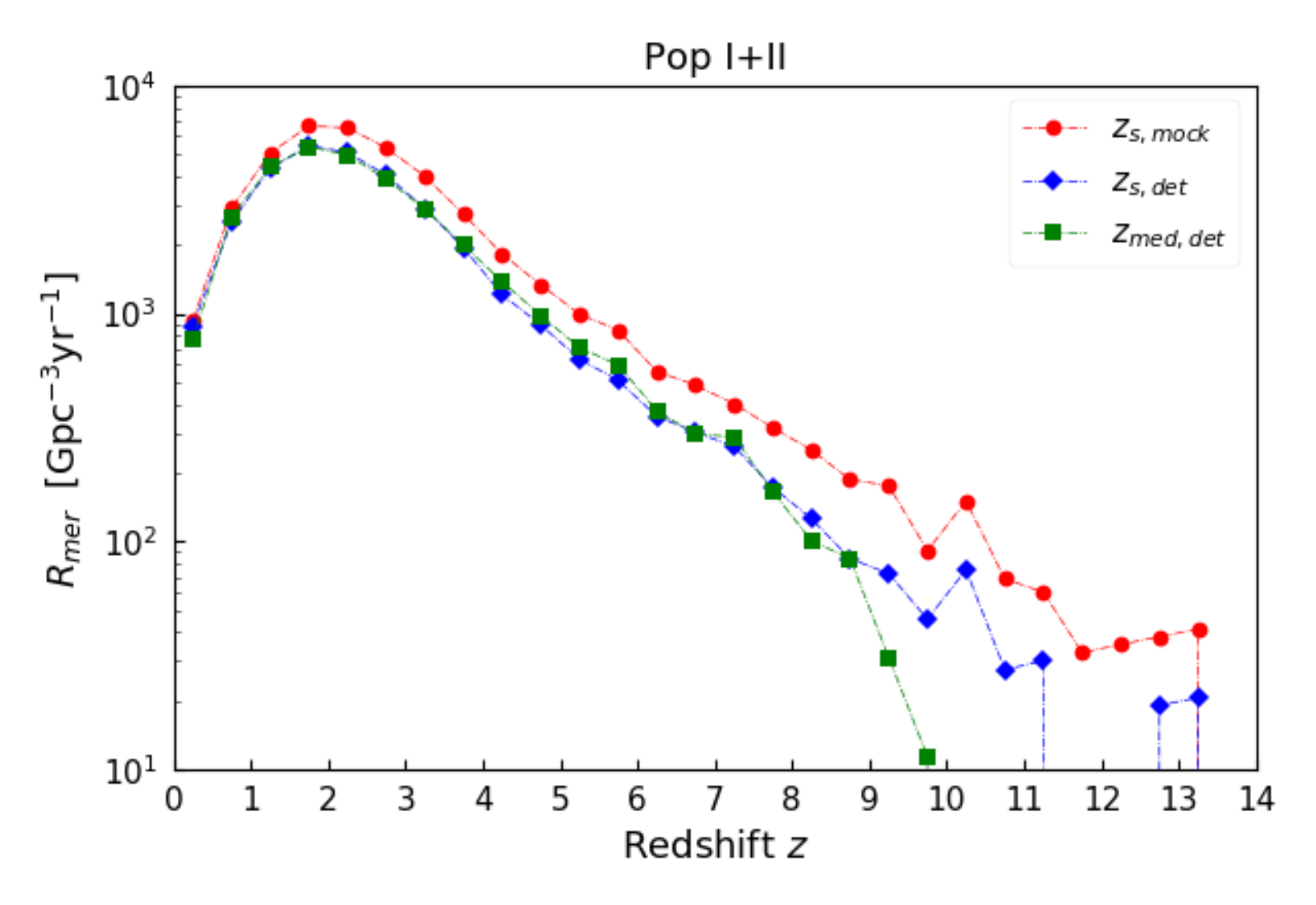}}\subfloat[\label{fig:rel_rate_m30}]{\includegraphics[width=0.9\columnwidth]{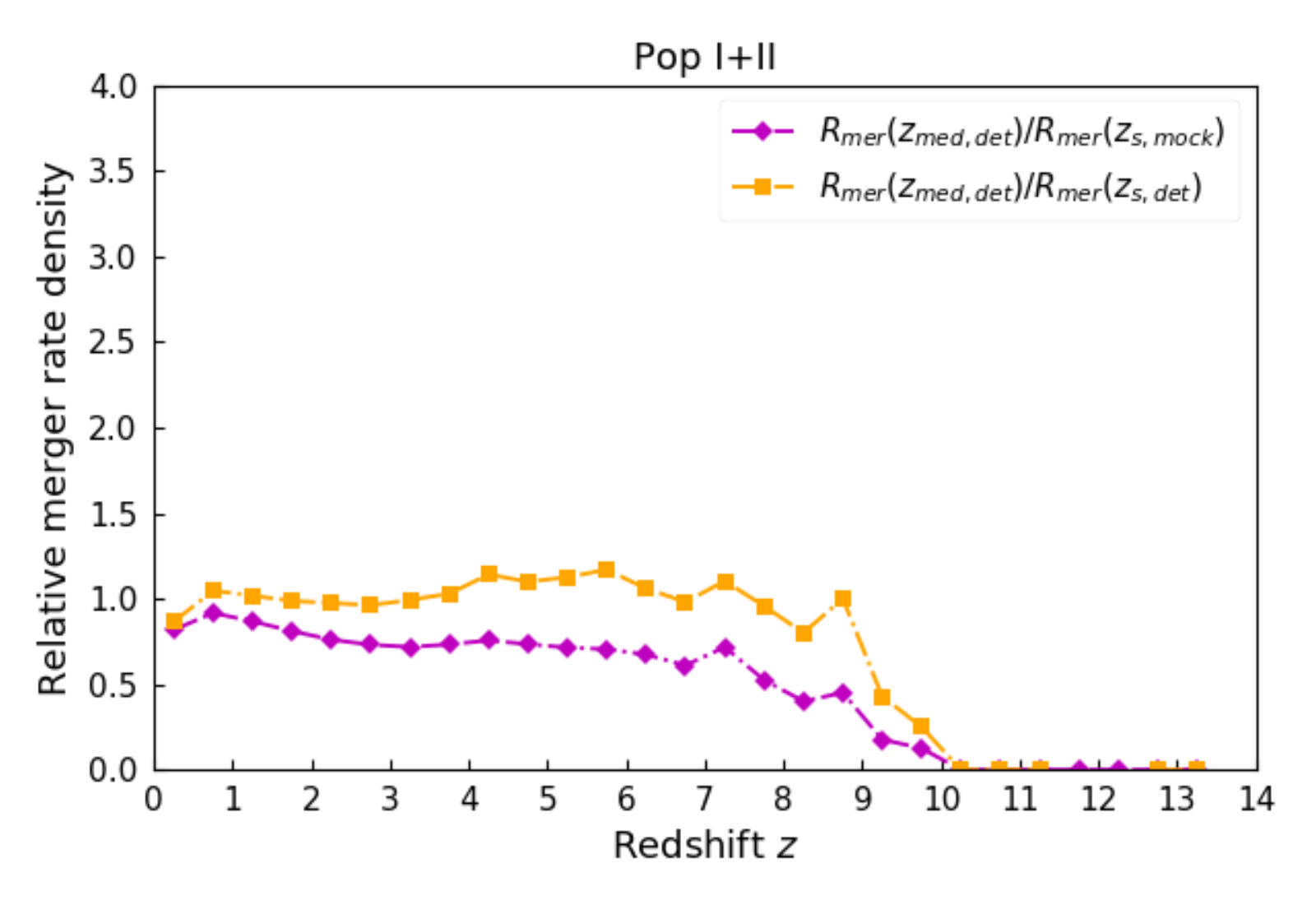}}\\
\subfloat[\label{fig:rate_pop3}]{\includegraphics[width=0.9\columnwidth]{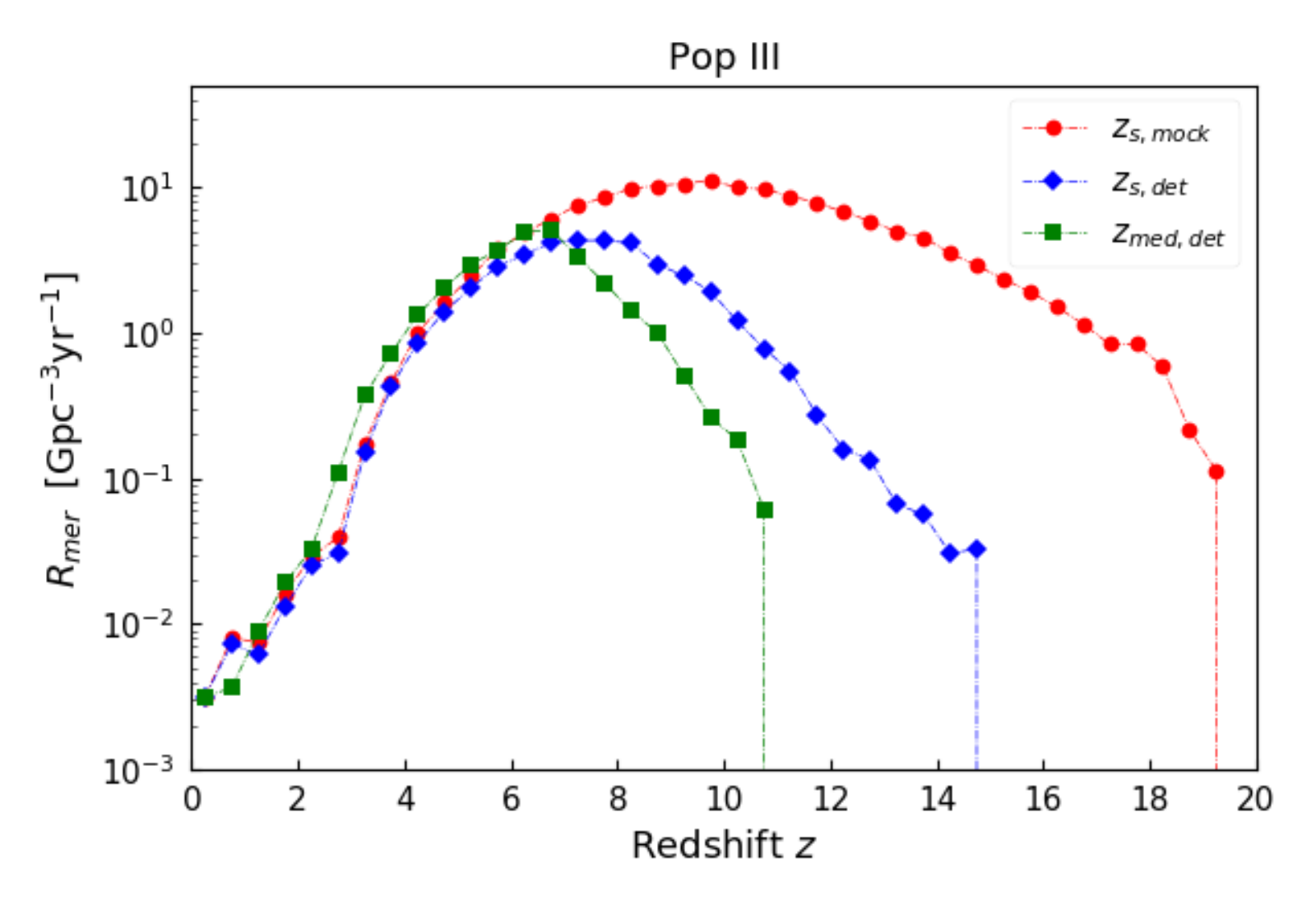}}\subfloat[\label{fig:rel_rate_pop3}]{\includegraphics[width=0.9\columnwidth]{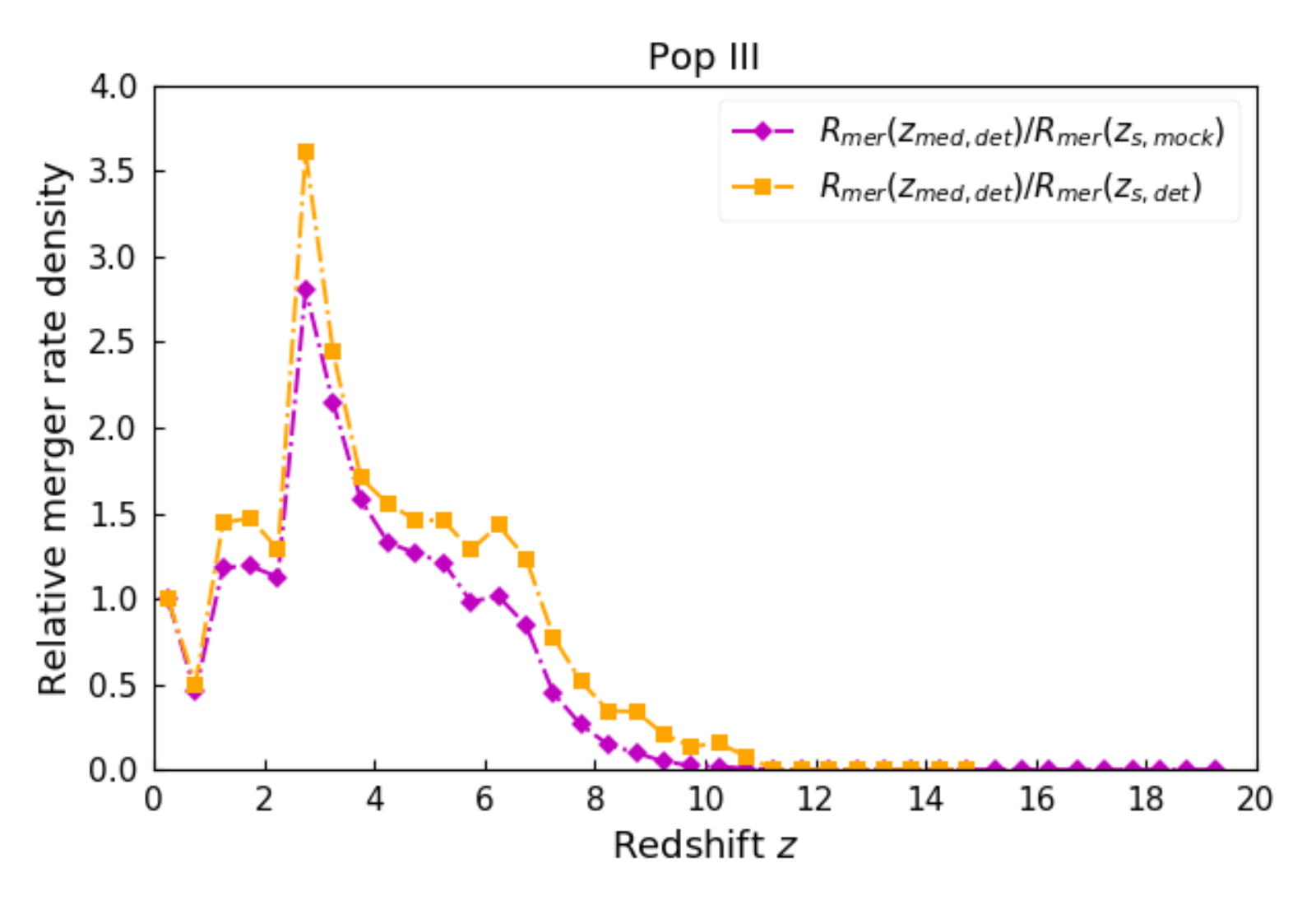}}\\
\subfloat[\label{fig:rate_cluster}]{\includegraphics[width=0.9\columnwidth]{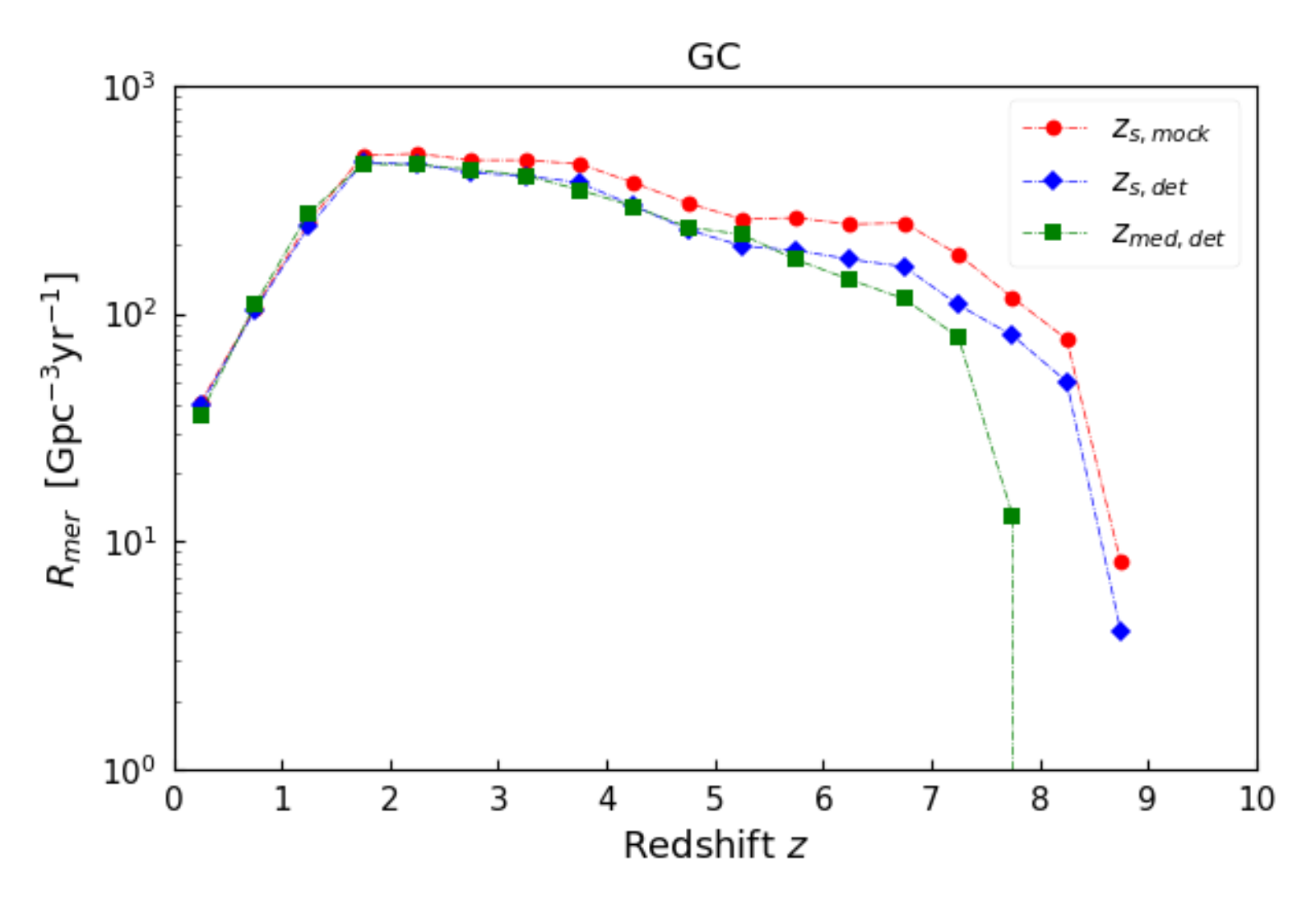}}\subfloat[\label{fig:rel_rate_cluster}]{\includegraphics[width=0.9\columnwidth]{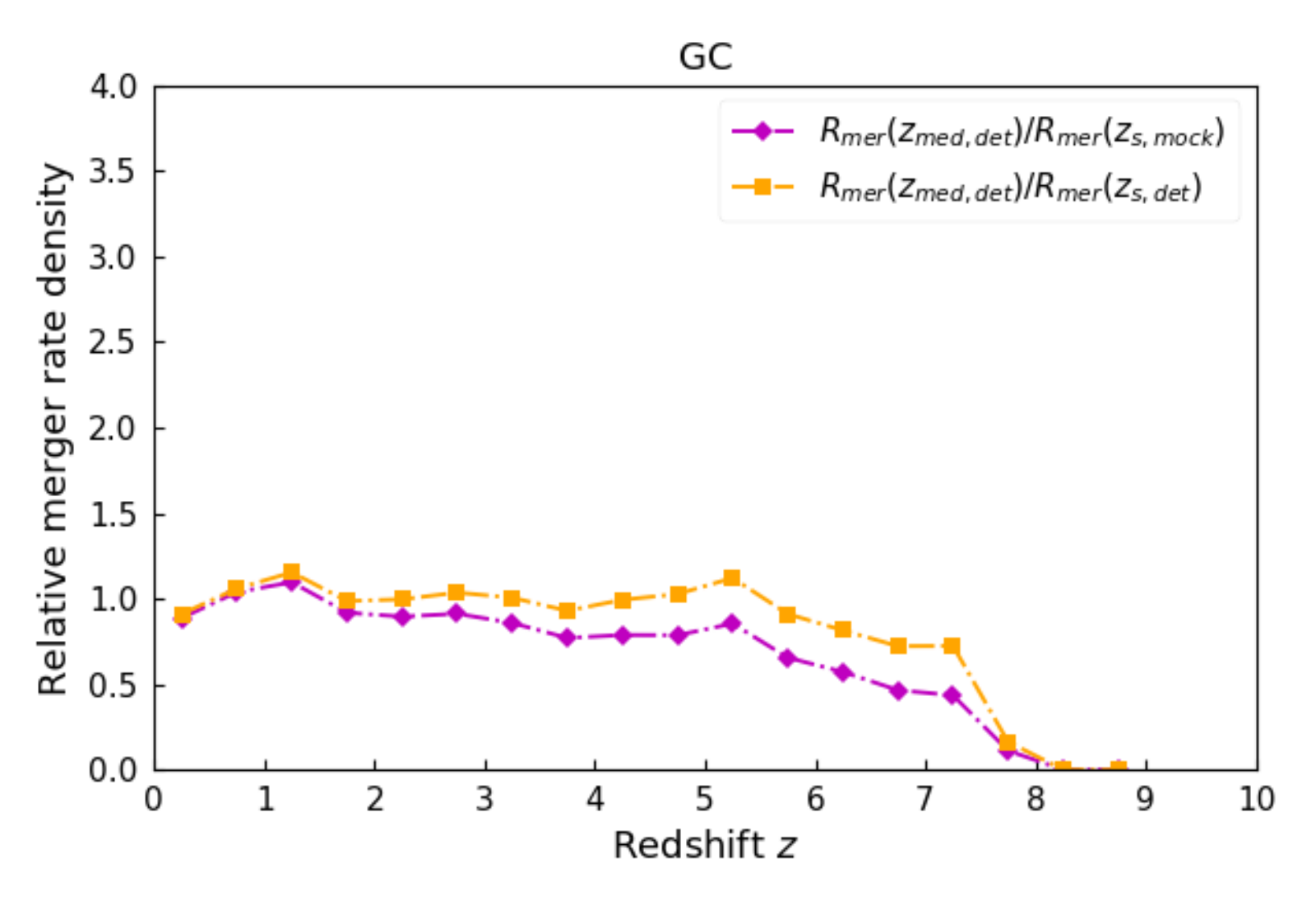}}\\
\caption{Merger rate density as a function of redshift. The left panel shows the merger rate density for the duration $T_{mock}$, calculated for every 0.5 redshift bin. The right plot shows the relative merger rate density. Top: Pop I+II. Middle: Pop III. Bottom: GC.}
\label{fig:merger_rate}
\end{figure*}

To generate the inspiraling compact binary sources originating from Pop I+II stars, we begin with the StarTrack model M30.B \citep{2020A&A...636A.104B}. While the code can give us a lot of information about the compact binaries, we use only the following data for each binary: (i) the masses for the merging compact objects $m^i_{1,2}$;  (ii) the delay time between the formation of the binary (ZAMS)  and its coalescence, $t_{del}^i$; and (iii) the metallicity of the binary $Z_i$.

A StarTrack population synthesis evolution \citep{2002ApJ...572..407B, 2008ApJS..174..223B} is performed by simulating a number of binaries with a given initial distribution of parameters. This corresponds to a simulation of $M_{sim}$ mass of stars, where one takes into account the binary fraction and the exact initial mass function for all stars and not just binaries. We simulate a fraction of the total population, defined by the condition $M_1> 6M_\odot$, at ZAMS, in order to include all stars that may form NSs or BHs over the course of their evolution, and to avoid simulating binaries that will not contribute to the formation of double compact objects. Then we assume  the standard distribution of binary parameters, namely the Kroupa initial mass function for the more massive component and the flat mass ratio distribution, and the 0.5 binary fraction. With these assumptions we calculate the $M_{sim}$, the total mass of all stars that must accompany the stellar evolution, leading to formation of compact object binaries. These include the binaries with $M_1<6 M_\odot$, as well as single stars.  The SFR history is assumed following \citep{2017ApJ...840...39M}: 

\begin{equation}\label{sfr}
    SFR(z) = 0.01\left(\frac{(1+z)^{2.6}}{1+((1+z)/3.2)^{6.2}}\right) M_{\odot}\rm{Mpc^{-3} yr^{-1}}
\end{equation}

To  generate the inspiraling compact binary sources originating for Pop III stars, we used the SFR given by \citep{2011A&A...533A..32D}. The population of stars known as Pop III.2 are those zero-metallicity primordial stars that form in an initially ionised gas \citep{2006MNRAS.366..247J, 2007ApJ...663..687Y}. We use the SFR calculated for these Pop III.2, assuming the velocity of the metal-enriched galactic wind propagating outwards from a central galaxy to be $v = 50$ km/s, representing low chemical enrichment \citep{2011A&A...533A..32D}.

\cite{2013MNRAS.432.3250K} found that most of the GCs now present formed during two distinct epochs of redshift, $z\sim2-3$ and $z\gtrsim6,$ with $\sim50\%$ of the total population having estimated ages of $\gtrsim12$ Gyr. They explored different assumptions on the minimum mass of the haloes in which proto-GC systems form. This corresponds to different assumptions on the minimum virial temperature $T_{vir}$ of the host haloes in which GC systems are allowed to form. We use the GC formation rate, with the assumption of $T_{vir} > 1.5 \times 10^5$K, as shown in Fig. 6 in \cite{2013MNRAS.432.3250K}.

Figure \ref{fig:sfr_pop} shows the SFR as a function of redshift for the three population classes that we used to generate our mock populations. The red curve shows the SFR for Pop I+II, obtained from \citep{2017ApJ...840...39M}. The SFR for Pop III obtained from \citep{2011A&A...533A..32D} is shown in blue, and the SFR for GC formation shown in green is from  \citep{2013MNRAS.432.3250K}. 

Additionally, we have to take into account the evolution of metallicity with redshift, which is given in \citep{2017ApJ...840...39M} as:

\begin{equation}\label{zzz}
    Z = 0.02\times 10^{0.153-0.074z^{1.34}}.
\end{equation}
For the purpose of this calculation, we make a non-physical assumption that there is a one to one dependence of the metallicity on redshift.  The function given in Equation (\ref{zzz}) can be easily inverted to obtain

\begin{equation}\label{ZZZ}
    z=\left[\frac{-\log(Z/0.02)+0.153}{0.074} \right]^{1/1.34}.
\end{equation}
We use Equation (\ref{ZZZ})  to calculate the redshift intervals that correspond to a given metallicity. We  assume that the stars are formed with this metallicity in a given redshift interval. For the given grid, we assume that the metallicity was constant and equal to $Z_i$ in the redshift interval of $z_{i-1}$ to $z_i$, as defined in \S \ref{pop_description} in Equations (\ref{Zi_m30}), (\ref{Zi_pop3}), and (\ref{Zi_cluster}). Thus, the metallicity dependent star formation rate is

\begin{equation}
SFR(z,Z)= \left\{
\begin{array}{ll}
SFR(z) & {\rm if \ \ \ } z_{i-1}(Z)<z<z_i(Z) \\
0 & {\rm otherwise} 
\end{array}  \right..
\end{equation}

The simulations provide us with N binaries for which we have their masses $m^i_{1,2}$, delay times $t_{del}^i$, and metallicity $Z_i$.  A binary (ZAMS) is formed at cosmic time $t_{ini}$, corresponding to redshift $z_{ini}=z(t_{ini})$. It evolves for $t_{evol}$, and merges so that $t_{del} = t_{evol} + t_{merg}$, where $t_{evol}$ denotes the time from ZAMS to the compact binary formation. This corresponds to the redshift $z_{del}=z(t_{del})$. The merger rate density per unit redshift of given binary $i$ from the simulation, as a function of redshift, is given by 

\iffalse
\begin{equation}\label{Rofi}
\mathcal{R}_i(z) = \frac{1}{(1+z)}\frac{dV}{dz}\left(\frac{SFR(z(t(z_{ini})+t_{evol}),Z)}{M_{sim}}\right)
\end{equation}
\fi
\begin{equation}\label{Rofi}
\mathcal{R_i}(z) = \frac{1}{(1+z)}\frac{dV}{dz}\left(\frac{SFR(z_{i, ini}),Z_i)}{M_{sim}}\right).
\end{equation}
The values of $M_{sim}$ for PopI+II, PopIII, and GC are $2.8\times 10^8 M_{\odot}$, $3.5 \times 10^9 M_{\odot}$ and $2.82 \times 10^8 M_{\odot}$, respectively. The definition of $M_{sim}$, the numbers involved in calculating $M_{sim}$, and the value of $M_{sim}$ for each of the three populations are given in: (i) PopI+PopII: Sect. A.7 of the Appendix, paragraph 4 in \cite{2020A&A...636A.104B}; (ii) PopIII: Table 4 in \cite{2017MNRAS.471.4702B}; and (iii) GC: Sect. 4 in \citep{2017MNRAS.464L..36A}. The details of assumed initial primary mass function, mass ratio, and initial separation of those binary stars for the three populations are given in: (i) PopI+PopII: Sect. A.7 of the appendix in \cite{2020A&A...636A.104B}; (ii) PopIII: Table 2 in \cite{2017MNRAS.471.4702B}; and (iii) GC: Table 1 (and the accompanying note) in \citep{2017MNRAS.464L..36A}.

We use the above as the redshift dependence of the merger rate density of each binary and notice that the probability density of a merger of a type $i$ to happen in the  Universe at redshift $z$ is proportional to $\mathcal{R}_i(z)$:

\begin{equation}\label{probzi}
P(i,z) \propto \mathcal{R}_i(z),
\end{equation}
which is  discreet in the index $i$ and continuous in z.  The probability distribution can be obtained by normalisation.

We draw random binaries from the distribution given by Equation (\ref{probzi}) and obtain a synthetic population of mergers. Each compact binary system drawn in this manner is then assigned random values to the four angles: the angle of declination $\delta$, the right ascension $\alpha$, the polarisation angle $\psi,$ and the inclination angle $\iota$ of the binary with respect to the direction of observation. We choose $\cos\delta, \alpha/ \pi$, $\cos \iota$ and $\psi/ \pi$, to be uncorrelated and distributed uniformly over the range $[-1,1]$.

\section{Results}\label{results}

Using the models for Pop I+II, Pop III, and GC compact binary object populations, we generate the mock populations for our analysis. For all the three sets of mock compact binary population, we simulated $N_{mock}$, the number of binaries. Out of these binaries, $N_{det}$ are detected, based on the chosen detection threshold. Then the time during which these binaries in the mock population are expected to merger is 
\begin{equation}
     T_{mock} = N_{mock}/N_{yr} \; \rm{yr},
\end{equation}
where $N_{yr}$ is the expected number of mergers per year calculated by integrating the merger rate density given by Equation (\ref{Rofi}). 

In order to estimate the constraints on the chirp mass $\mathcal{M}$ and redshift $z$, we follow the algorithm described in detail in SB2. Thus, we get the probability distribution for chirp mass and redshift for each detected source. The median values of each of these estimated distributions of chirp mass and redshift for each detected compact binary source, $\mathcal{M}_{med,det}$ and $z_{med, det}$, are then compared with the actual source parameters, $\mathcal{M}_{s}$ and $z_{s}$.

The merger rate density $R_{mer}$ for each of the three mock populations is calculated for every redshift bin of 0.5 for the time $T_{mock}$ of the simulation of the whole set of compact binary sources as:

\begin{equation}\label{merger_rate_den}
    R_{mer}(z_i, z_{i+1}) = \frac{1+z_{i+1}}{\int^{z_{i+1}}_{z_i} \frac{dV}{dz}dz}\times \frac{N_{z_i:z_{i+1}}}{T_{mock}},
\end{equation}
where $N_{z_i:z_{i+1}}$ is the number of mergers in a redshift bin of $(z_i:z_{i+1})$. 

\subsection{Summary of the mock populations}

In the mock population representing the population of compact binaries originating from Pop I+II stars, we generate a set of 25862 sources. This corresponds to a time period of $T_{mock} = 0.006$ years. The redshift of these sources range from $z_{min} = 0.03$ to  $z_{max} = 13.04$, and the chirp mass is within the range $\mathcal{M}_{min} = 0.96M_{\odot}$ to $\mathcal{M}_{max} = 48.47M_{\odot}$.

From the population model of compact binaries from Pop III stars only, we obtain 23990 sources within the redshift range of $z_{min} = 0.28$ to  $z_{max} = 19.36$, and the chirp mass range from $\mathcal{M}_{min} = 2.27M_{\odot}$ to $\mathcal{M}_{max} = 40.04 M_{\odot}$. For this set of mock population, $T_{mock} = 9.878$yr.

We generate 26461 compact binary sources from the GC binary population corresponding to $T_{mock} = 0.073$yr with the redshift range $z_{min} = 0.06$ to  $z_{max} = 8.63$, and the chirp mass range from $\mathcal{M}_{min} = 2.77 M_{\odot}$ to $\mathcal{M}_{max} = 554.63M_{\odot}$. The details of all the three mock populations are summarised in Table \ref{tab:pop_details}.

\subsection{Mass and redshift estimates}

We now compare the three sets of parameters: (i) the actual parameters of the sources in the mock population, $\mathcal{M}_{s, mock}$ and $z_{s, mock}$; (ii) the actual parameters of the binaries that were detected ,$\mathcal{M}_{s, det}$ and $z_{s, det}$; and (iii) the estimated medians of the parameters for each of the detected compact binaries, $\mathcal{M}_{med, det}$ and $z_{med, det}$.

The cumulative distribution of the parameters in the three populations are shown in Figure \ref{fig:cummulative_all}. The top panel shows the Pop I+II binaries, the middle panel shows the Pop III compact binary sources, and the bottom panel shows the distributions for the GC population of compact binaries. The dot-dashed red curve represents the distribution of the parameter value in a given mock population. The dashed blue curve represents the distribution of the parameter value of the detected binary systems out of a given mock population. The solid green curve represents the distribution of the estimated median of the parameter value.

The distribution for the parameters for the compact binaries originating from Pop I+II stars is shown in Figures \ref{fig:chm_m30_cummu} and \ref{fig:red_m30_cummu}. It shows that the binaries in this mock population are located up to $z_{s, mock} \sim 13$, and this whole range of redshift is detectable. The estimated redshift of the detected Pop I+II binaries sources is $z_{med, det} \lesssim 10$. This shows that we underestimate the redshift for these binaries. Looking at the estimates of the chirp mass in Figure \ref{fig:chm_m30_cummu}, we see that the whole range of chirp masses is detectable, but we overestimate the range of $\mathcal{M}$ for this population of binaries to $\mathcal{M}_{med, det} \sim 55$.

In the case of Pop III binaries, shown in Figures \ref{fig:chm_pop3_cummu} and \ref{fig:red_pop3_cummu}, the mock population has sources with $\mathcal{M}_{s, mock} \lesssim 40 M_{\odot}$ located up to redshift $z_{s, mock} \sim 20$. The whole mass range of sources is detected for $z_{s, det} \lesssim 15 $. We largely overestimate the range of chirp mass and underestimate the redshift of these detected sources to $\mathcal{M}_{med, det} \lesssim 52 $ and $z_{med, det} \lesssim 11 $. 

Figures \ref{fig:chm_cluster_cummu} and \ref{fig:red_cluster_cummu} show the distribution of parameters for the GC binaries. The distribution shows that the binaries over the whole range of redshift of $z_{s, mock} \sim 9$ are detectable, but only masses with $\mathcal{M}_{s , det} \lesssim 215 M_{\odot}$ are detectable, and  $\sim 99\% $ of the detected sources have $\mathcal{M}_{s, det} \lesssim 50 M_{\odot}$. The chirp mass of the binaries of the GC population is slightly overestimated, while the redshift is slightly underestimated.

\begin{figure*}
         \centering
         \includegraphics[scale = 1]{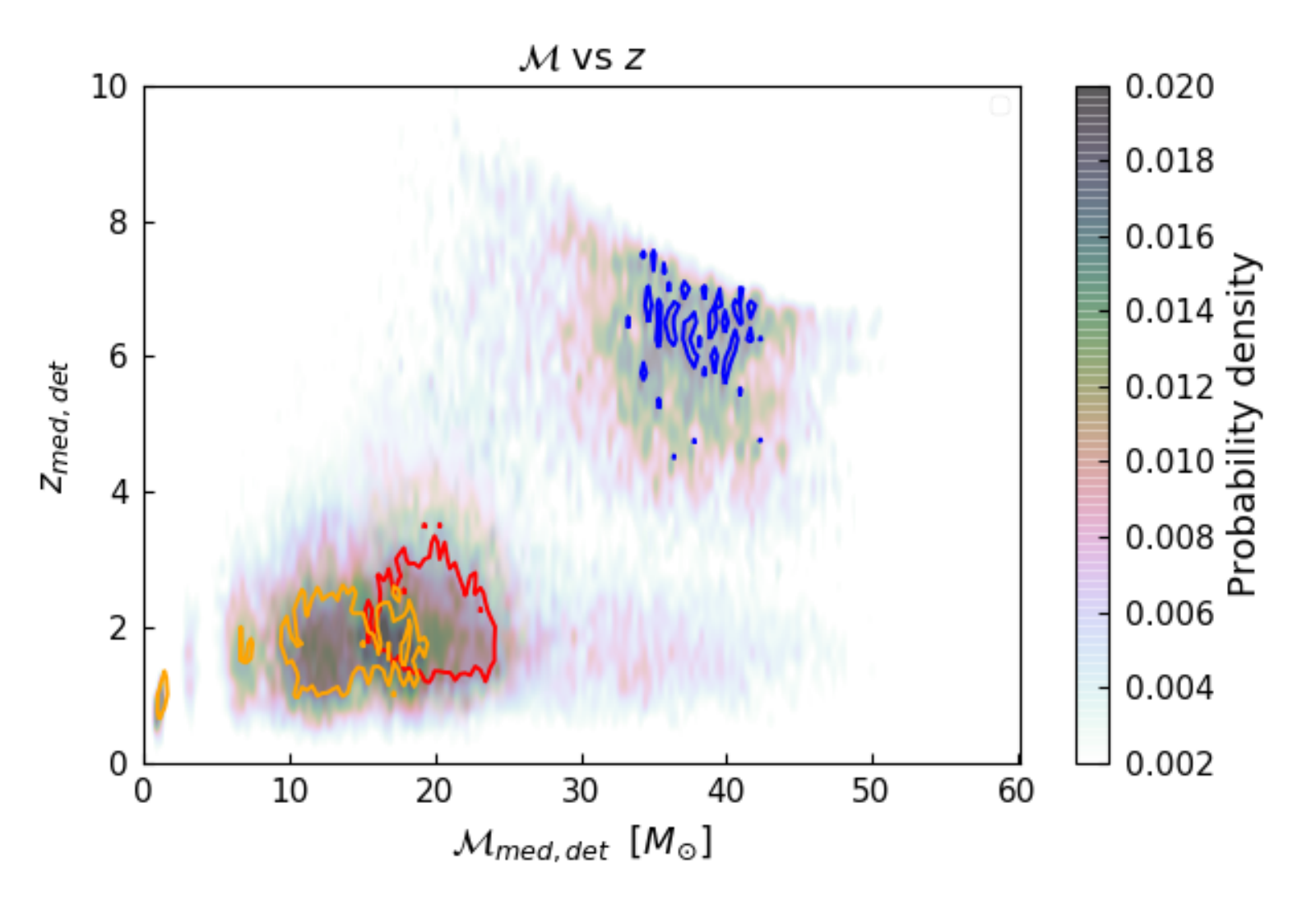}
\caption{Pop I+II, Pop III, and the GC population are plotted together with the contours showing the 90\% probability region of the estimated medians of $\mathcal{M}$ and $z$, for the three set of populations. The orange, blue, and red contours enclose the 90\% probability region of the estimated median values of Pop I+II, Pop III, and the GC population, respectively. The density value shown in the colour bar corresponds to the density values for each of the three sets of the populations normalised individually. }
\label{fig:combined_pop}
\end{figure*}

\begin{figure*}
         \centering
         \includegraphics[scale = 1]{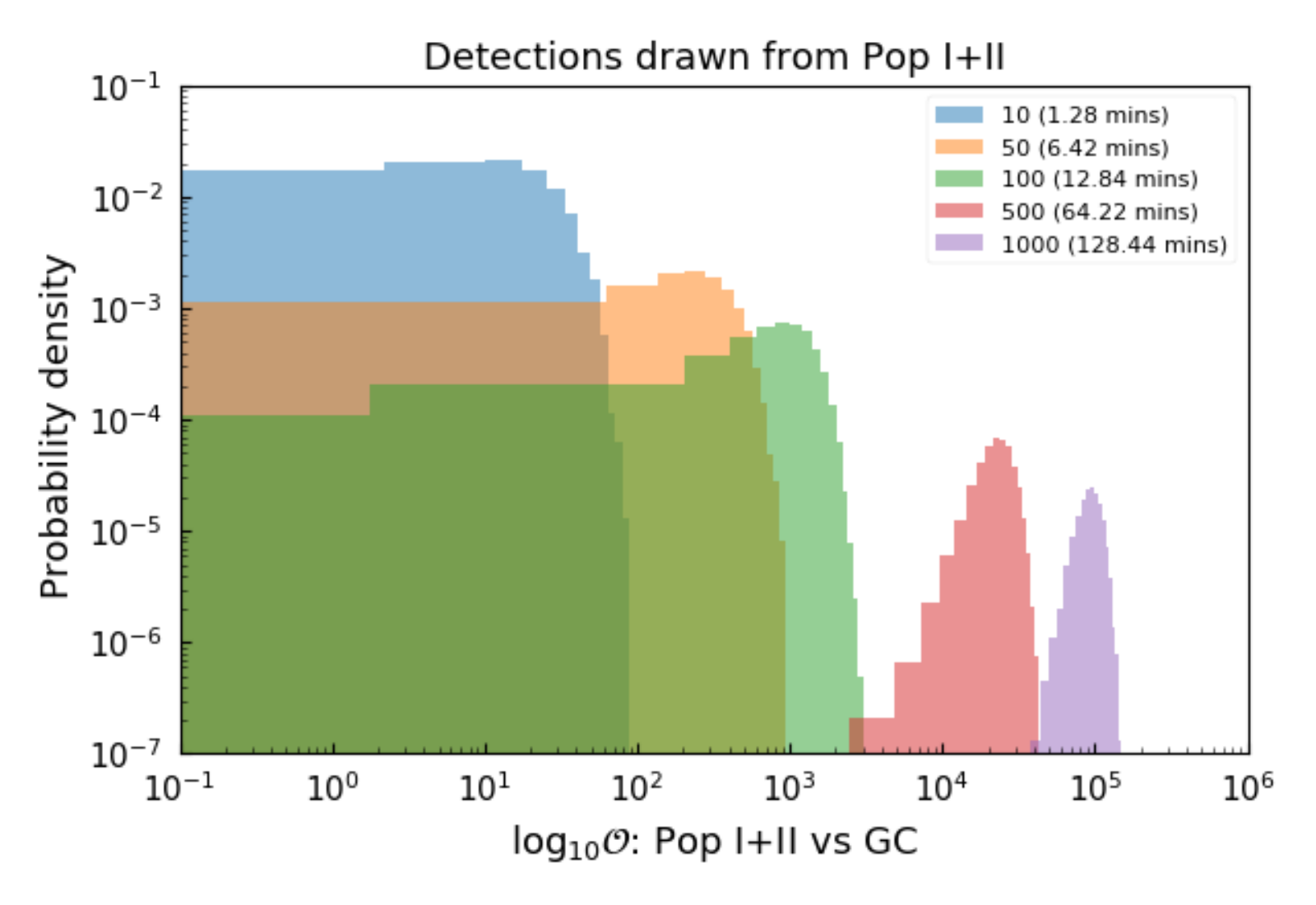}
\caption{Odds ratio $\mathcal{O}$ for Pop I+II vs the GC population. We draw a random set of detections (10, 50, 100, 500, and 1000) originating from Pop I+II. The values mentioned in the brackets are the duration of the observations for the corresponding number of detections.}
\label{fig:bayes}
\end{figure*}

\begin{figure*}
\centering
\subfloat[\label{fig:KL_m30}]{\includegraphics[width= 0.9\columnwidth]{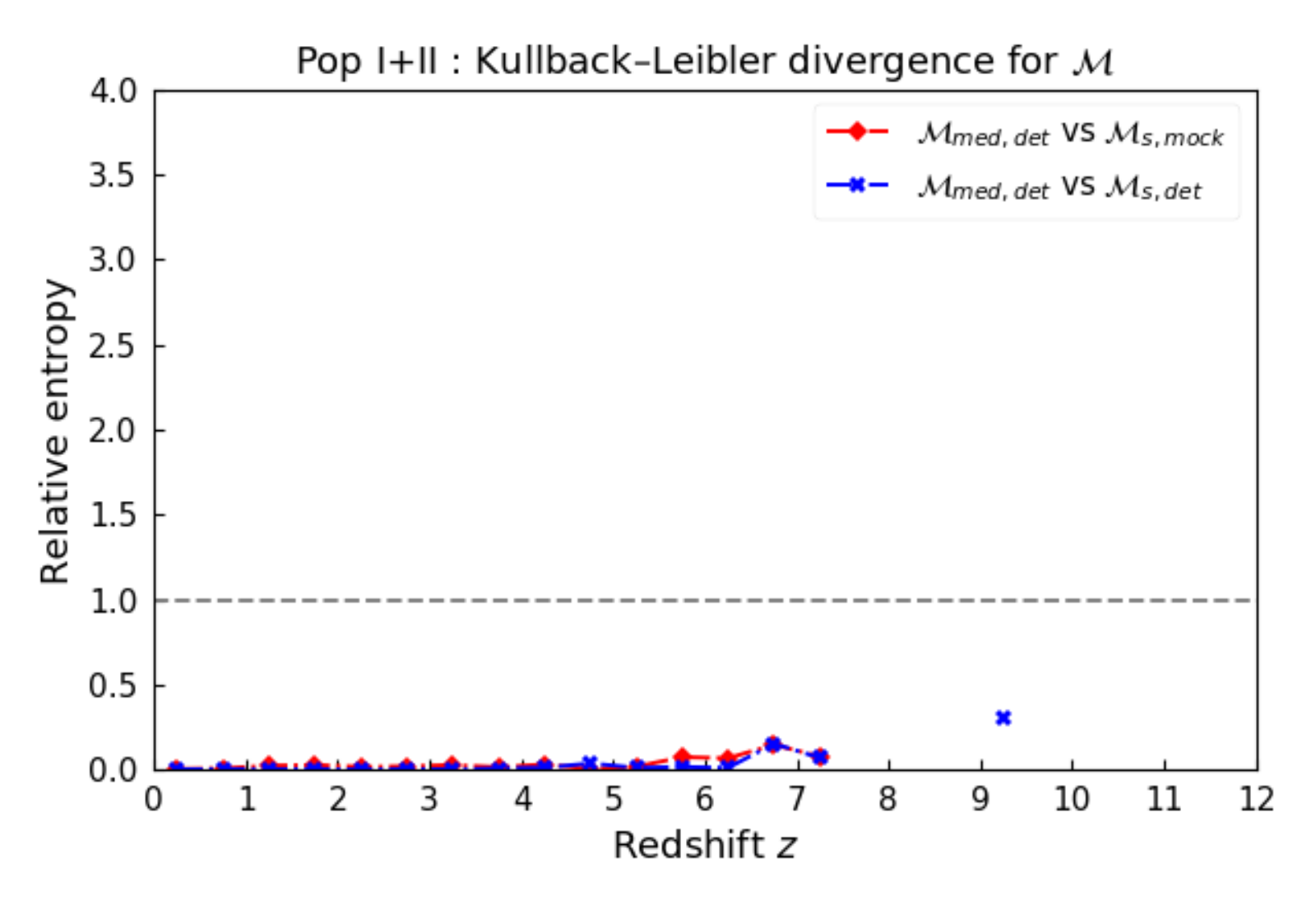}}
\subfloat[\label{fig:KL_pop3}]{\includegraphics[width=0.9\columnwidth]{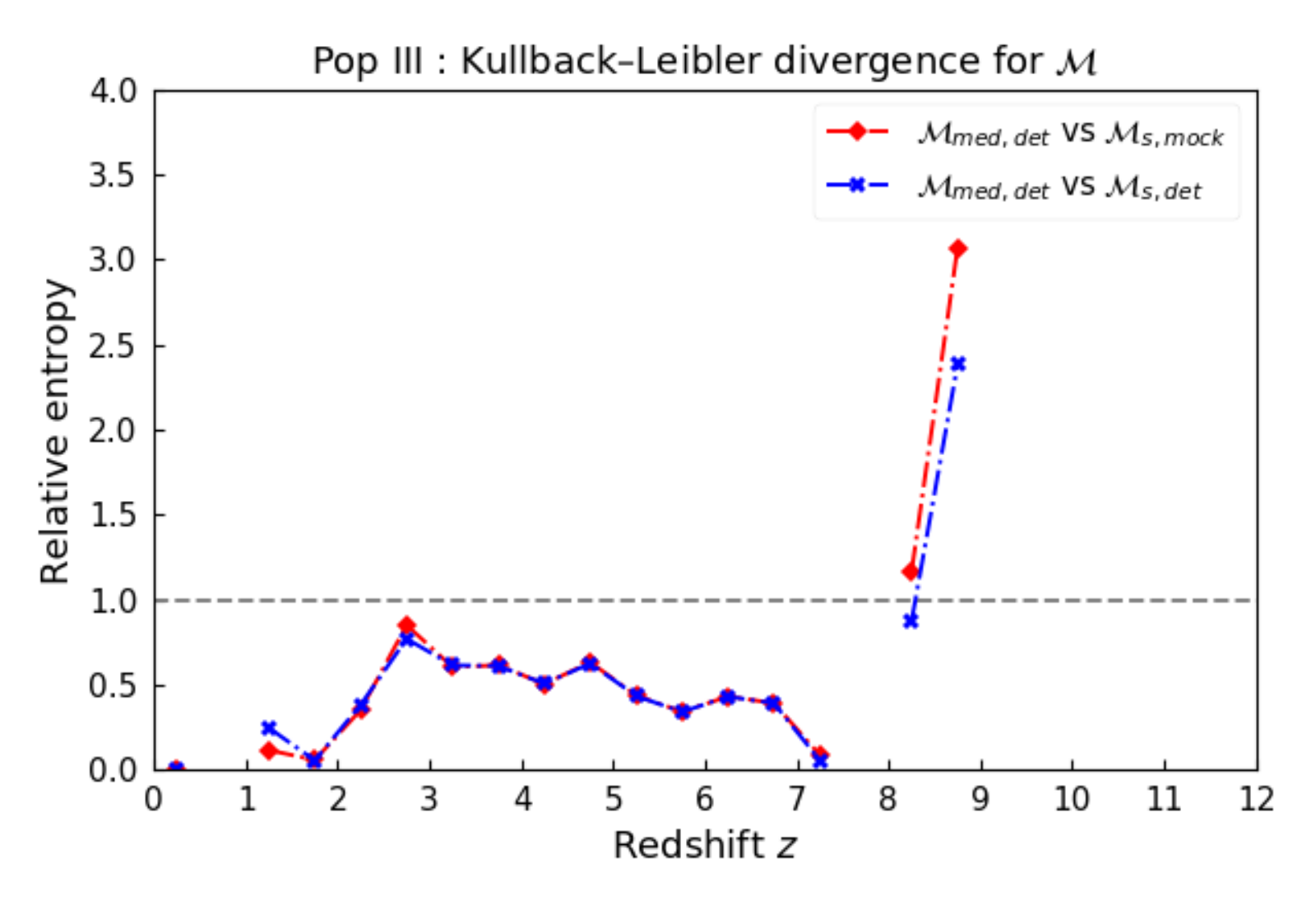}}\\
\subfloat[\label{fig:KL_cluster}]{\includegraphics[width=0.9\columnwidth]{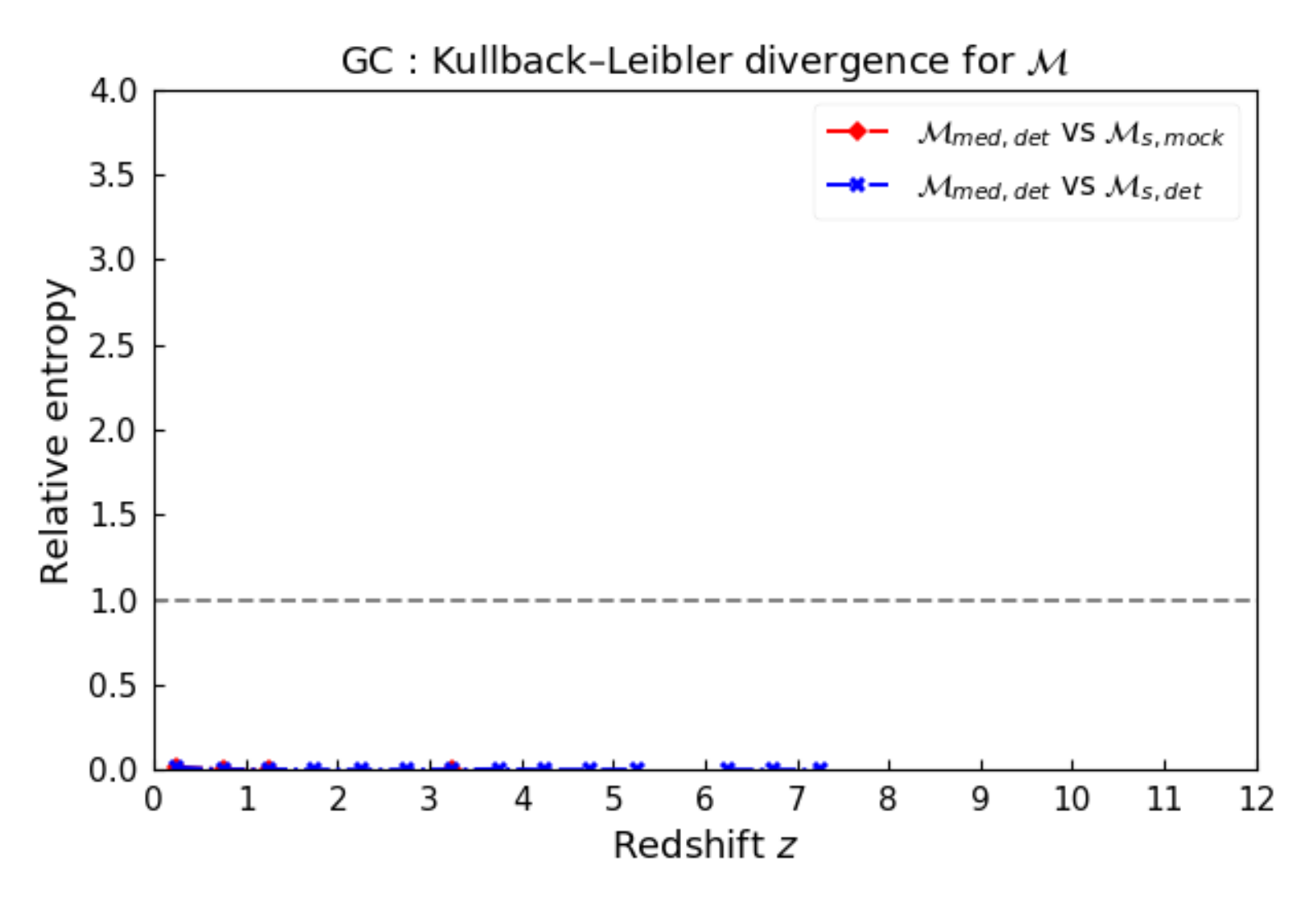}}
\caption{Kullback–Leibler divergence for the chirp mass, calculated for the binary sources of Pop I+II,  Pop III, and the GC populations in a redshift bin of 0.5. The grey line denotes our assumed threshold value of $D_{KL}(p||q) = 1$.}
\label{Kl_test}
\end{figure*}

We now continue to compare the parameters of the detected binary sources. Figure \ref{fig:injvsmedian} shows the 2D probability density distribution  for the comparison between the estimated median values ($\mathcal{M}_{median, det},z_{median, det}$) with respect to the actual values of the parameters ($\mathcal{M}_{s,det},z_{s, det}$),  for each detected compact binary source in a given population. The blue contour encloses the 90\% probability region of all the detected sources. The total number of detected sources in each mock population $N_{det}$ is mentioned in Table \ref{tab:pop_details}. Chirp mass is shown in the left panel and redshift in the right panel.

Figures \ref{fig:chm_m30} and \ref{fig:red_m30} show the probability density distribution $\mathcal{M}$ and $z$ for the detected compact binaries from Pop I+II stars. We see that for 90\% of the detected sources, which lie within $\mathcal{M}_{s, det} \lesssim 20 M_{\odot}$ and $z_{s, det} \lesssim 4$, we have $\mathcal{M}_{median, det}\approx \mathcal{M}_{s,det}$ and $ z_{median, det} \approx z_{s, det}$.  The probability density peaks of $\mathcal{M}_{s, det}$ of the detected compact binaries in Pop I+II population are mainly concentrated in five mass regions below $20 M_{\odot}$ and are estimated correctly.

Figures \ref{fig:chm_pop3} and \ref{fig:red_pop3} show the probability density distribution $\mathcal{M}$ and $z$ for detected compact binaries from Pop III stars. We see the two probability density peaks of $\mathcal{M}_{s, det}$ are correctly estimated but for 90\% of the detected sources, which lie within $ 35 \lesssim \mathcal{M}_{s, det}/M_{\odot} \lesssim 38 $ and $ 3.5 \lesssim z_{s, det} \lesssim 10 $, we have $\mathcal{M}_{median, det} \gg \mathcal{M}_{s,det}$ and $ z_{median, det} \ll z_{s, det}$. Since most of the detected binaries are located at $ 3.5 \lesssim z_{s, det} \lesssim 10 $, these generate a low S/N, and therefore they have higher errors in the estimation of the parameters.

Figures \ref{fig:chm_cluster} and \ref{fig:red_cluster} show the probability density distribution $\mathcal{M}$ and $z$ for detected compact binaries from stars evolving in the GC. We see that regions of the five probability density peaks of $\mathcal{M}_{s, det}$ are correctly estimated. For 90\% of the detected sources, which lie within $ 11 \lesssim \mathcal{M}/M_{\odot} \lesssim 25 $ and $ 0.5 \lesssim z \lesssim 3.5 $, we have $\mathcal{M}_{median, det} \gtrsim \mathcal{M}_{s,det}$ and $ z_{median, det} \lesssim z_{s, det}$. It should be noted that the chirp mass distributions for the GC data we used are sensitive to the prescriptions for evolution of BH progenitors. With different prescriptions, such as rapid or delayed supernovae, or winds, these values could be higher and perhaps that might lead to better detectability and recovery of parameters.

We now consider the picture of the analysis in the $(\mathcal{M}-z)$ space, for the three sets of parameters: (i) the actual parameters of the sources in the mock population, $\mathcal{M}_{s, mock}$ and $z_{s, mock}$; (ii) the actual parameters of the binaries that were detected, $\mathcal{M}_{s, det}$ and $z_{s, det}$; and (iii) the estimated medians of the parameters for each of the detected compact binaries, $\mathcal{M}_{med, det}$ and $z_{med, det}$. Figures \ref{fig:chm_z_m30}, \ref{fig:chm_z_pop3}, and \ref{fig:chm_z_cluster}, show the 2D probability densities of the chirp mass and redshift of the compact binary systems for Pop I+II, Pop III, and the GC populations, respectively.

In all the three figures, the top panel shows the parameters $\mathcal{M}_{s, mock}$ and $z_{s, mock}$, the middle panel shows the probability density of $\mathcal{M}_{s, det}$ and $z_{s, det}$, while the bottom panel shows the probability density the parameters $\mathcal{M}_{med, det}$ and $z_{s, det}$.

In Figure \ref{fig:chm_z_m30} we see that the probability density peaks in $\mathcal{M}-z$ are clearly visible in all three panels for the three sets of parameters, but most low-mass sources $\mathcal{M} \lesssim 5 M_{\odot}$ are not detected beyond redshift $z \approx 6, $ and the redshift of the lighter binaries is underestimated. The blue contour encloses the region of 90\% probability for the estimated parameters $\mathcal{M}_{med, det}$ and $z_{med, det}$ of all the 20528 detected binaries.

Figure \ref{fig:chm_z_pop3} shows the 2D probability density in the $(\mathcal{M}-z)$ space for the Pop III compact binary sources. The probability density of $\mathcal{M}-z$ is seen to peak in two close regions in the limit $ 35 \lesssim \mathcal{M}_{s, mock}/M_{\odot} \lesssim 38 $  for $ 2 \lesssim z \lesssim 20$ and these compact binaries are detectable within a redshift range of $2 \lesssim z \lesssim 12$. As noted earlier, the low S/N generated by these sources results in a larger error on the estimated parameter values. Due to this, the redshift is largely underestimated, while the chirp mass is largely overestimated. We can thus see that the features of mass distributions for the Pop III binary systems are not estimated accurately and a refined population model should be considered for further analysis. The blue contour encloses the region of 90\% probability for the estimated parameters $\mathcal{M}_{med, det}$ and $z_{med, det}$ of all the 9683 detected binaries.

The 2D probability density in the $(\mathcal{M}-z)$ space for the GC compact binary sources is shown in Figure \ref{fig:chm_z_cluster}. Most of the detectable lighter binaries with $\mathcal{M}_{s, det} \lesssim 5 M_{\odot}$ are within $z \approx 7$. There is an underestimation of the redshift of the binaries detected beyond $z \approx 1$, as seen earlier in Figure \ref{fig:red_cluster}, which results in the overestimation of the chirp mass of these binaries. The blue contour encloses the region of 90\% probability for the estimated parameters $\mathcal{M}_{med, det}$ and $z_{med, det}$ of all the 22937 detected binaries.

\subsection{Recovery of merger rate history}

We estimate the merger rate density $R_{mer}(z)$ for the time duration  $T_{mock}$ for every 0.5 redshift bin using Equation (\ref{merger_rate_den}). We bin the redshift values of all the binary sources in a given mock population $z_{s, mock}$ to obtain the number of binaries in the given redshift bin, and then using Equation (\ref{merger_rate_den}), obtain $R_{mer}(z_{{(s, mock)}_i}, z_{{(s, mock)}_{i+1}})$. Thus, we get $R_{mer}(z_{s, mock})$. The merger rates $R_{mer}(z_{s, det})$ and $R_{mer}(z_{med, det})$ are calculated similarly. We also calculate the relative merger rate densities $\frac{R_{mer}(z_{med, det})}{R_{mer}(z_{s, mock})}$ and $\frac{R_{mer}(z_{med, det})}{R_{mer}(z_{s, det})}$ for comparison.

The results are shown in the Figure \ref{fig:merger_rate}. The left panel shows $R_{mer}(z)$ and the right panel shows the corresponding relative merger rate densities. The top, middle, and bottom panels show the estimates for Pop I+II binaries, Pop III, and the GC population of compact binaries.

Figures \ref{fig:rate_m30} and \ref{fig:rel_rate_m30} show the rate estimates for the Pop I+II compact binaries. $R_{mer}(z_{s, det})$ and $R_{mer}(z_{med, det})$ are in good agreement for $z \lesssim 8.75 $. For larger redshifts the deviation tends to increase.

The merger rate estimates for Pop III compact binary systems are shown in Figures \ref{fig:rate_pop3} and \ref{fig:rel_rate_pop3}. The large underestimation of higher redshift values results in $R_{mer}(z_{med, det}) \gg R_{mer}(z_{s, det})$ for $ 2 \lesssim z \lesssim 6$ and $R_{mer}(z_{med, det}) \ll R_{mer}(z_{s, det})$ for $ z \gtrsim 7$.

Figures \ref{fig:rate_cluster} and \ref{fig:rel_rate_cluster} show the rate estimates for the GC compact binaries.
$R_{mer}(z_{s, det})$ and $R_{mer}(z_{med, det})$ are in good agreement for $z \lesssim 7.5$. At higher redshifts, $R_{mer}(z_{med, det}) \ll R_{mer}(z_{s, det})$.

\subsection{Distinguishability of populations in the $\mathcal{M}-z$ plane}

We analysed the parameters such as chirp mass $\mathcal{M}$, redshift $z$, and the merger rate of the inspiraling compact binary systems of Pop I+II, Pop III, and the GC population using a single ET. In order to see the distinguishability of each population, we plot all of them in Figure \ref{fig:combined_pop}. The normalisation of each of the three sets depends on the merger rates of each of these populations. We generate the 2D density for each set and normalise them individually. We calculate the 90\% probability region for $\mathcal{M}-z$ values in each of these three sets of populations.

In Figure \ref{fig:combined_pop}, the orange contours enclose the 90\% probability region of the estimated medians of the parameters of Pop I+II compact binaries. The blue contours enclose the 90\% probability region of the estimated medians of the parameters of the compact binaries for the Pop III and the red contours enclose the 90\% probability region of the estimated medians of the parameters of the compact binaries for the GC. The three regions are clearly distinguishable.

In order to further investigate the distinguishability of compact binaries, we calculate the odds ratio. We do so in the following way for two population models A (Pop I+II) and B (GC). We generate the observed  population from  model A, with the detected parameters $M_i, z_i$, where $i=1,N$ and $N$ is the number of detections.  We calculate the probability density of detection in the $\mathcal{M}-z$ plane for each of the two models A and B,  $\frac{dP_{A}}{dz dM}$ and $\frac{dP_{B}}{dz dM} $, and we calculate the ratio of seeing the detected population, assuming these calculated probability densities. Thus we have\begin{equation}\label{odds_ratio}
O_{AB}= \frac{{\Pi_i \frac{dP_{A}}{dz dM}(z_i, M_i) }}{{{\Pi_i \frac{dP_{B}}{dz dM}(z_i, M_i) }}},
\end{equation}
where $i$ denotes all the measured values of parameters. In this calculation we conservatively neglect the detected points for which the detection probability is zero.

To calculate the odds ratio of the Pop I+II and the GC population, we draw a random set of detections (10, 50, 100, 500, and 1000) originating from Pop I+II and calculate the odds ratio $\mathcal{O}$ for Pop I+II vs the GC population using Equation (\ref{odds_ratio}). Each of these set of detections are drawn $10^4$ times. The priors on Pop I+II and the GC population were assumed to be the same. Figure \ref{fig:bayes} shows the distribution of the $\rm{log}_{10}\mathcal{O}$ for Pop I+II vs the GC population. It can be seen that, even for a small number of 500 detections, which are observed within $\sim 1$ hr, the two populations can be clearly distinguished.

\subsection{Recovery of mass function}

    Each population of compact binary system is characterised by a specific $\mathcal{M}-z$ distribution. Although the full space of these parameters for each set of populations has been shown, we now proceed with the analysis of the estimated distributions of chirp mass values. As was mentioned earlier, the left panel in Figure \ref{fig:cummulative_all} shows the cumulative distribution of $\mathcal{M}$ in the three populations. The dot-dashed red curve represents the distribution of the parameter value in a given mock population, the dashed blue curve represents the distribution of the parameter value of the detected binary systems out of a given mock population, and the solid green curve represents the distribution of the estimated median of the parameter value. In order to closely investigate these distributions of $\mathcal{M}$ further, we analyse in every redshift bin of 0.5. We compare the values of $\mathcal{M}_{med, det}$ and $\mathcal{M}_{s, det}$, and also compare $\mathcal{M}_{med, det}$ and $\mathcal{M}_{s, mock}$ by calculating the relative entropy or Kullback–Leibler divergence ($D_{KL}$) \citep{cover1999elements, kullback1951information,2014arXiv1404.2000S} to see the recovery of the features in chirp mass distribution for each of the three mock population sets. Relative entropy or $D_{KL}$ is defined as the measure which quantifies how close a given probability distribution $p = {p_i}$ is to a model distribution $q = {q_i}$. In other words, the relative entropy $D_{KL}(p||q)$ is a measure of the inaccuracy of the assumption that the distribution is $q$ when the true distribution is $p$:

\begin{equation}
   D_{KL}(p||q) = \sum_{i} p_i \log \left( \frac{p_i}{q_i}\right).
\end{equation}
Relative entropy is always non-negative, and is zero if and only if $p=q$ \citep{cover1999elements}. The comparisons for $\mathcal{M}_{s, det}$ with $\mathcal{M}_{med, det}$, and $\mathcal{M}_{s, mock}$ with $\mathcal{M}_{med, det}$, for the three mock populations are shown in Figure\ref{Kl_test}. We assume that for $D_{KL}(p||q) > 1$, the distributions cannot be considered to be similar. 

Figure \ref{fig:KL_m30}, shows the $D_{KL}$ as a function of redshift for the compact binary sources originating from Pop I+II stars, in each redshift bin of 0.5. We see that $D_{KL}(\mathcal{M}_{s, mock}||\mathcal{M}_{med, det}) \approx 0$ and $D_{KL}(\mathcal{M}_{s, det}||\mathcal{M}_{med, det}) \approx 0$ for $z \approx 6,$ and is $<1$ for the full redshift range. Thus, we conclude that we recover the chirp mass characteristics for this mock population correctly. 

For the mock population of Pop III compact binaries, $D_{KL}$ as a function of redshift is shown in Figure \ref{fig:KL_pop3}. $0< D_{KL}(\mathcal{M}_{s, mock}||\mathcal{M}_{med, det}) < 1$ and $0 < D_{KL}(\mathcal{M}_{s, det}||\mathcal{M}_{med, det}) < 1$ for $ 1 < z < 7.5$.  At larger redshifts, we do not recover the chirp mass distribution correctly, as discussed previously, due to the overestimation of redshift values, as shown in Figure \ref{fig:red_pop3}.

In the case of the GC compact binaries, shown in Figure \ref{fig:KL_cluster}, we see that 
$D_{KL}(\mathcal{M}_{s, mock}||\mathcal{M}_{med, det}) \approx 0$ and $D_{KL}(\mathcal{M}_{s, det}||\mathcal{M}_{med, det}) \approx 0$ for the full redshift range. We therefore conclude that we recover the chirp mass characteristics for this mock population correctly as well.

\section{Conclusion}\label{conc}

We analysed the compact binary populations of (i) Pop I+II, (ii) Pop III, and (iii) GCs, with a single ET using the ET-D design sensitivity. The three sets of populations represent different metallicities, and hence different ages of the binary populations, and they represent different formation scenarios. Thus, the analysis presents the capability of a single ET to detect and distinguish different compact binary populations. We took into account the effect of rotation of the Earth on the antenna pattern and estimated the parameters chirp mass $\mathcal{M}$, redshift $z$, and merger rate $R_{mer}$ for each of the three sets of the populations. We show that if the populations of compact binaries are separated in $\mathcal{M}-z$ space, then the ET as a single instrument is capable of detecting and distinguishing these different compact binary populations. If the populations overlap in $\mathcal{M}-z$, then it will be necessary to consider other parameters such as spin to estimate the distinguishability.

We found that, based on the chosen detection threshold, 90\% of the detected  Pop I+II binaries lie within $z \lesssim 3.5$, Pop III up to $ z\lesssim 9 $ and GC binaries up to $z \lesssim 5$. The deviation of the estimated values of the chirp mass, redshift, and merger rate density, compared with the actual source values of these parameters for the detected binaries, is much larger in the case of Pop III than in the case of Pop I+II, and the GC binaries. Since the features of mass distributions for the Pop III binary systems are not estimated accurately, a detailed analysis with a refined population model should be taken into consideration  to recover the correct distribution for this population of compact binaries. Assuming that a sufficient number of binaries are detected from each of these populations,  we show that the three sets of populations are distinguishable in the combined bulk detection. 

We checked the accuracy of the recovery of the chirp mass for each redshift bin of 0.5 in each of the population sets by calculating the Kullback–Leibler divergence. For the Pop I+II  and the GC compact binary sources, we recover the chirp mass characteristics correctly. However, in the case Pop III population, due to the overestimation chirp mass of sources at large redshift value of $z > 7.5$, we do not recover the chirp mass characteristics correctly beyond this redshift. We thus find that that the mass distribution characteristics of different compact binary populations can be estimated with a single ET.

\begin{acknowledgements}
%\section{Acknowledgement}\label{sec:Acknow}
NS and TB are thankful to Prof. Mirek Giersz for permitting the use of MOCCA data. We thank the reviewer for all valuable comments and suggestions which helped us to improve the manuscript. NS and TB acknowledge the support from the Foundation for Polish Science grant TEAM/2016-3/19 and NCN grant UMO-2017/26/M/ST9/00978. NS is supported by the "Agence Nationale de la Recherche”, grant n. ANR-19-CE31-0005-01 (PI: F. Calore). AA acknowledges support from the Swedish Research Council through the grant 2017-04217. This document has been assigned Virgo document number VIR-0785A-21, and ET document number ET-0010A-22.
\end{acknowledgements}

\bibliography{reference}
\bibliographystyle{aa}

\end{document}